\documentclass[fleqn,usenatbib]{mnras}

\usepackage{mathptmx}

\usepackage[T1]{fontenc}
\usepackage[utf8]{inputenc}
\usepackage{ae,aecompl}

\usepackage{graphicx}	
\usepackage{amsmath}	
\usepackage{amssymb}	

\newcommand{\msun}{$\; \rm M_\odot$}

\usepackage{url,times,graphicx,hyperref,amsmath,amsfonts,amssymb,aas_macros,color,epsfig,epstopdf}
\usepackage{booktabs}

\newcommand{\Mpc}            {\,{\rm Mpc}}
\newcommand{\kms}            {\,{\rm km\,s^{-1}}}
\newcommand{\vmax}           {V_{\rm max}}
\newcommand{\Msun}           {\rm M_{\odot}}

\definecolor{purple}{rgb}{0.5, 0., 0.5}

\newcommand{\red}{\textcolor{red}}

\title[Magellanic satellites in $\Lambda$CDM]{Magellanic satellites in
  $\Lambda$CDM cosmological hydrodynamical simulations of the Local Group}

\author[Santos-Santos et al.]{
\parbox[t]{\textwidth}{
Isabel M.E. Santos-Santos$^{1}$\thanks{E-mail: isantos@uvic.ca}, Azadeh Fattahi$^{2}$, Laura V. Sales$^{3}$, Julio F. Navarro$^{1}$
}\\
\\
$^{1}$Department of Physics and Astronomy, University of Victoria, Victoria, BC V8P 5C2, Canada\\
$^{2}$Institute for Computational Cosmology, Department of Physics, Durham University, South Road, Durham, DH1 3LE, UK\\
$^{3}$Department of Physics and Astronomy, University of California Riverside, 900 University Avenue, CA 92507, USA\\
}

\pubyear{2020}

\begin{document}
\label{firstpage}
\pagerange{\pageref{firstpage}--\pageref{lastpage}}
\maketitle

\begin{abstract}
  We use the APOSTLE $\Lambda$CDM cosmological hydrodynamical
  simulations of the Local Group to study the recent accretion of
  massive satellites into the halo of Milky Way (MW)-sized galaxies. These
  systems are selected to be close analogues to the Large Magellanic
  Cloud (LMC), the most massive satellite of the MW. The
  simulations allow us to address, in a cosmological context, 
  the impact of the Clouds on the MW,
  including the contribution of Magellanic satellites to the MW
  satellite population, and the constraints placed on the Galactic
  potential by the motion of the LMC. We show that
  LMC-like satellites are twice more common around Local Group-like
  primaries than around isolated halos of similar mass; these
  satellites come from large turnaround radii 
  and are on highly eccentric orbits whose velocities at first
  pericentre are comparable with the primary's escape velocity. This
  implies $V_{\rm esc}^{\rm MW} (50 $ kpc$)\sim 365$ km/s, a strong
  constraint on Galactic potential models.  LMC analogues contribute
  about 2
  satellites with $M_*>10^5\, M_\odot$, having thus
  only a mild impact on the luminous satellite population of their
  hosts. 
  At first
  pericentre, LMC-associated satellites are close to the LMC in
  position and velocity, and are distributed along the LMC's orbital
  plane. Their orbital angular momenta roughly align with the
  LMC's, but, interestingly, they may appear to ``counter-rotate'' the
  MW in some cases. These criteria refine earlier estimates of the LMC
  association of MW satellites: 
  only the SMC, Hydrus1, Car3, Hor1, Tuc4, Ret2 and Phoenix2 
  are compatible with all
  criteria.  Carina, Grus2, Hor2 and Fornax are 
  less probable associates given their large LMC relative velocity.
\end{abstract}



\section{Introduction}
\label{SecIntro}

It is now widely agreed that the Large Magellanic Cloud (LMC),
the most luminous satellite of the Milky Way (MW), is at a particular
stage of its orbit. Its large Galactocentric velocity ($\sim 328$
km/s) is dominated by the tangential component ($\sim 320$ km/s) and
is much higher than all plausible estimates of the MW circular velocity
at its present distance of $\sim 50$ kpc \citep[see; e.g.,][and
references therein]{Kallivayalil2013,GaiaColl2018}. This implies that
the LMC is close to the pericentre of a highly eccentric orbit with
large apocentric distance and long orbital times. Together with the
presence of a clearly asociated close companion \citep[the Small
Magellanic Cloud, SMC, see; e.g.,][]{Westerlund1990,DOnghia2016}, the
evidence strongly suggests that the Clouds are just past their first
closest approach to the Galaxy
\citep{Besla2007,BoylanKolchin2011a,Patel2017}.

The particular kinematic stage of the LMC, together with the
relatively high stellar mass of the Clouds
\citep[$M_*\sim 2.5\times 10^9\, M_\odot$;][]{Kim1998}, offer clues
about the MW virial\footnote{We shall refer to the virial boundary of
  a system as the radius where the mean enclosed density is
  $200\times$ the critical density for closure. We shall refer to
  virial quantities with a ``200'' subscript.} mass and insight into
the hierarchical nature of galaxy clustering in the dwarf galaxy
regime.

Clues about the MW mass fall into two classes.  One concerns
the relation between virial mass and satellite statistics; namely, the
more massive the Milky Way halo the higher the likelihood of hosting a
satellite as massive as the LMC. Empirically, observational estimates
suggest that up to $\sim 40\%$ of $L^*$ galaxies may host a satellite
as luminous as the LMC within $\sim 250$ kpc and up to a $10\%$ chance
of having one within $\sim 50$ kpc \citep{Tollerud2011}. This result
has been interpreted as setting a lower limit on the MW virial mass
of roughly $\sim 10^{12}$ M$_\odot$
\citep{Busha2011,BoylanKolchin2011a,Patel2017, Shao2018}.

The other class relates to kinematics; if the LMC is near its first
pericentric passage, its velocity, not yet affected substantially by
dynamical friction, should reflect the total acceleration experienced
during its infall. If, as seems likely, that infall originated far from the MW virial
boundary, then the LMC velocity would provide a robust estimate of the
MW escape velocity at its present location.  This assumes, of course,
that the LMC is bound to the MW, an argument strongly supported by its
status as the most luminous and, hence, most massive
satellite. Unbound satellites are indeed possible, but they tend to
occur during the tidal disruption of groups of dwarfs, and to affect
{\it only} the least massive members of a group \citep[see;
e.g.,][]{Sales2007}.

A strong constraint on the MW escape velocity at $r\sim 50$ kpc,
$V^{\rm MW}_{\rm esc}$, could help to discriminate between competing
Galactic potential models by adding information at a distance where
other tracers are scarce and where commonly-used Galactic potential
models often disagree \citep[see; e.g.,][]{Irrgang2013,
  Bovy2015,GaravitoCamargo2019,Errani2020}. For example,
$V^{\rm MW}_{\rm esc}$ at $50$ kpc vary between $\sim 450$ km/s and
$\sim 330$ km/s for the four Galactic models proposed in these
references.

The peculiar kinematic state of the LMC adds complexity to the
problem, but also offers unique opportunities. On the one hand, the
short-lived nature of a first pericentric passage implies that the MW
satellite population is in a transient state and out of
dynamical equilibrium. This compromises the use of simple equilibrium
equations to interpret the dynamics of the MW satellites, and reduces
the usefulness of the MW satellites as a template against which the
satellite populations of external galaxies may be contrasted.

However, it also offers a unique opportunity to study the satellites
of the LMC itself. If on first approach, most LMC-associated
dwarfs should still lie close to the LMC itself, as the Galactic tidal
field would not have had time yet to disperse them
\citep{Sales2011}. If we can disentangle the LMC satellite population
from that of the MW then we can directly study the satellite
population of a dwarf galaxy, with important applications to our ideas
of hierarchical galaxy formation \citep{DOnghia2008} and to the
relation between galaxy stellar mass and halo mass at the faint-end of
the galaxy luminosity function \citep{Sales2013}.

The issue of which MW satellites are ``Magellanic'' in origin has
been the subject of several recent studies, mainly predicated
on the idea that LMC satellites should today have positions and
velocities consistent with what is expected for the tidal debris of the
LMC halo \citep{Sales2011,Yozin2015,Jethwa2016}. One application of these
ideas is that LMC satellites should accompany the LMC orbital motion
and, therefore, should have orbital angular momenta roughly parallel
to that of the LMC.

Using such dynamical premises, current estimates
based on accurate proper motions from \textit{Gaia}-DR2 have suggested at least
four ultrafaint dwarfs (Car 2, Car 3, Hor 1, and Hydrus 1) as highly probable
members of the LMC satellite system
\citep{Kallivayalil2018,Fritz2018}, an argument supported and extended
further by semianalytic modeling of the ultrafaint population
\citep{Dooley2017b, Nadler2019}.

Taking into account the combined gravitational potential of the MW+LMC
system might bring two extra candidates (Phx 2 and Ret 2) into plausible
association with the LMC \citep{Erkal2020,Patel2020}. Revised kinematics
for the classical dwarfs have also led to suggestions that the Carina
and Fornax dSph could have been brought in together with the LMC
\citep{Jahn2019,Pardy2020}. Further progress requires refining and
extending membership criteria in order to establish the identity of the
true Magellanic satellites beyond doubt.

Much of the progress reported above has been made possible by LMC
models based on tailored simulations where the Milky Way and the LMC
are considered in isolation, or on dark matter-only cosmological
simulations where luminous satellites are not explicitly
followed. This paper aims at making progress on these issues by
studying the properties of satellite systems analogous to the LMC
identified in cosmological hydrodynamical simulations of Local Group
environments from the APOSTLE project.

The paper is organized as follows. We describe our numerical datasets
in Sec.~\ref{SecNumSims}, and the identification of LMC analogues in
APOSTLE in Sec.~\ref{SecLMCanalogs}. The satellites of such analogues,
and their effect on the primary satellite population, are explored in
Sec.~\ref{SecLMCAssocSats}. Finally, Sec.~\ref{SecLMCIdCrit} uses
these results to help identify Magellanic satellites in the MW and
Sec.~\ref{SecVesc} considers the constraints placed by the LMC on the
MW escape velocity and Galactic potential. We conclude with a brief
summary in Sec.~\ref{SecConc}.

\begin{figure*}
\includegraphics[width=0.7\linewidth]{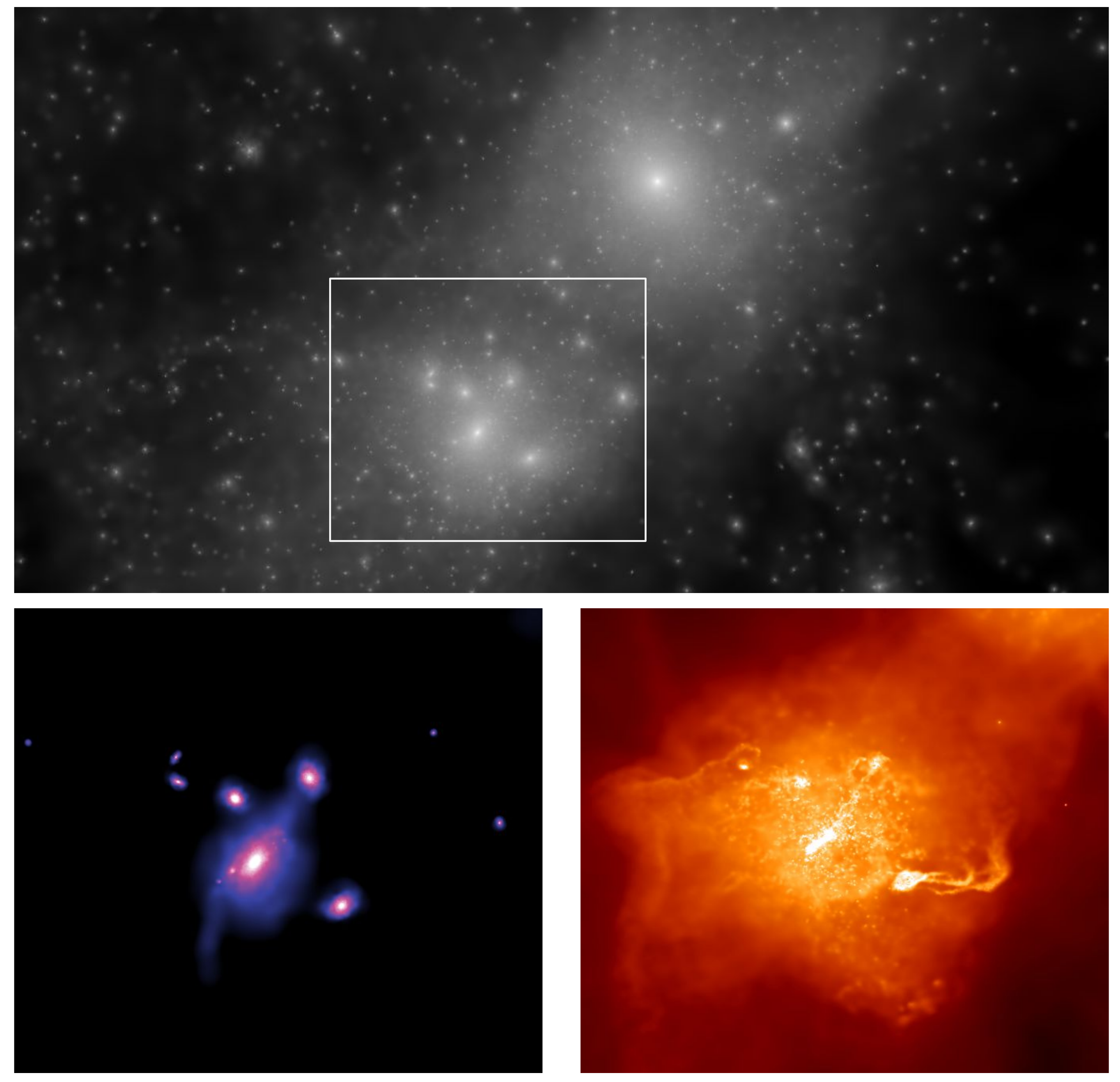}
\caption{ An image of an APOSTLE simulation volume that includes an
  LMC analogue as defined in this work (labelled 1-1-1 in subsequent
  figures and tables).  The upper panel shows the dark matter
  distribution of the Local group-like environment, with the M31
  analogue in the upper right part of the panel, and the MW analogue in
  the bottom left. The area enclosed in a rectangle, which includes
  the MW and LMC analogues, is shown in the bottom-left and bottom-right
  panels in stellar and gas density projections, respectively.  The
  LMC analogue is the object located on the lower right in the bottom
  panels.  Note the purely gaseous stream that emerges from it, with
  no stellar counterpart, reminiscent of a ``Magellanic stream''.  }
 \label{fig:LMCimage}
\end{figure*}


\section{Numerical Simulations}
\label{SecNumSims}

All simulations used in this paper adopt a flat $\Lambda$CDM model
with parameters based on WMAP-7 \citep{Komatsu2011}:
$\Omega_{\rm m}=0.272$, $\Omega_{\Lambda}=0.728$,
$\Omega_{\rm bar}=0.0455$, $H_0=100\, h\, \kms \, {\rm Mpc}^{-1}$,
$\sigma=0.81$, with $h=0.704$.

\subsection{The DOVE simulation}

We use the DOVE cosmological N-body simulation to study the frequency
of massive satellites around Milky Way-mass haloes and possible
environmental effects in Local Group volumes. DOVE evolves a
$100^3 \Mpc^3$ cosmological box with periodic boundary conditions
\citep{Jenkins2013} with $1620^3$ collisionless particles with mass
per particle $m_{\rm p}= 8.8 \times 10^6$ \msun. The initial
conditions for the box were made using \textsc{panphasia}
\citep{Jenkins2013} at $z=127$, and were evolved to $z=0$ using the
Tree-PM code \textsc{P-Gadget3}, a modified version of the publicly
available \textsc{Gadget-2} \citep{Springel2005b}.

\subsection{The APOSTLE simulations}

The APOSTLE project is a suite of ``zoom-in'' cosmological
hydrodynamical simulations of twelve Local Group-like environments,
selected from the DOVE box \citep[][]{Sawala2016}. These Local Group
volumes are defined by the presence of a pair of haloes whose masses,
relative radial and tangential velocities, and surrounding Hubble flow
match those of the Milky Way-Andromeda pair \citep[see][for
details]{Fattahi2016}.

APOSTLE volumes have been run with the EAGLE (Evolution and Assembly
of GaLaxies and their Environments) galaxy formation code
\citep{Schaye2015,Crain2015}, which is a modified
version of the Tree-PM SPH code {\sc P-Gadget3}. The subgrid physics
model includes radiative cooling, star formation in regions denser
than a metallicity-dependent density threshold, stellar winds and
supernovae feedback, homogeneous X-ray/UV background radiation, as
well as supermassive black-hole growth and active galactic nuclei
(AGN) feedback (the latter has substantive effects only on very
massive galaxies and its effects are thus essentially negligible in
APOSTLE volumes).

The model was calibrated to approximate the stellar mass function of
galaxies at $z=0.1$ in the stellar mass range of
$M_{\rm star}= 10^8$-$10^{12}\, M_\odot$, and to yield realistic
galaxy sizes.  This calibration means that simulated galaxies follow
fairly well the abundance-matching
relation of \citet{Behroozi2013} or \citet{Moster2013}
\citep[see][]{Schaye2015}.

Although dwarf galaxy sizes were not used to adjust the model,
  they are nevertheless in fairly good agreement with observational
  data \citep{Campbell2017}.  Isolated dwarf galaxies follow
  as well a tight $M_{\rm star}$-$V_{\rm max}$ relation \citep[see
  Fig.~1 in][]{Fattahi2018}, consistent with extrapolations of
  abundance-matching models.  The APOSTLE simulations have been run
at three different levels of resolution, all using the "Reference"
parameters of the EAGLE model. In this work we use the medium
resolution runs (labelled ``AP-L2''), with initial dark matter and gas
particle masses of $m_{\rm dm}\sim 5.9 \times 10^6 \Msun$ and
$m_{\rm gas}\sim 1.2\times10^5 \Msun$, respectively. As in DOVE,
haloes and subhaloes in APOSTLE are identified using a
friends-of-friends groupfinding algorithm \citep{Davis1985} and
\textsc{subfind} \citep{Springel2001}.  These have been linked between
snapshots by means of merger trees, which allow us to trace individual
systems back in time \citep{Qu2017}.

 \begin{figure}
\includegraphics[width=\columnwidth]{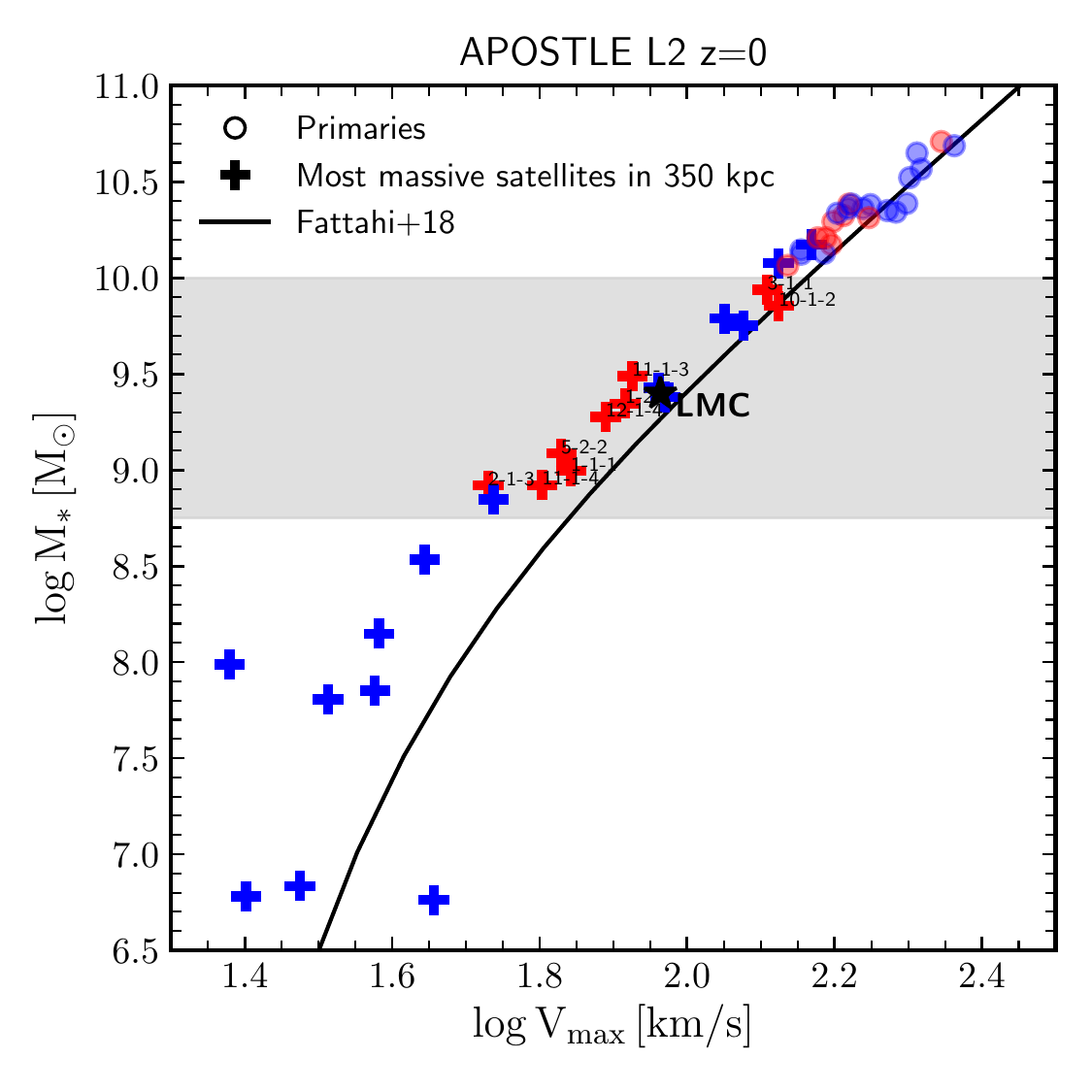}
\caption{$V_{\rm max}$-$M_{*}$ relation for the most massive
  satellites (crosses) of the 24 primaries (circles; i.e., MW and M31
  analogues) from the 12 APOSTLE-L2 volumes at $z=0$.  The shaded area delimits
  the $M_{*}$ range around the LMC's observed stellar mass value (star
  symbol) chosen to search for LMC-analogue candidates. The final LMC
  analogues that were selected for analysis in this work (see
  Sec.~\ref{SecLMCOrbits}), and their corresponding primaries, are
  shown in red.  A line shows the average
  $V_{\rm max}$-$M_{*}$ relation for APOSTLE centrals from
  \citet{Fattahi2018}. }
\label{fig:vmaxmstar}
\end{figure}

\subsection{Galaxy identification}
\label{SecGalID}

Particles in the simulations are grouped together using the
friends-of-friends algorithm \citep[FoF;][]{Davis1985}, with a linking
length of $0.2$ times the mean inter-particle separation. Self-bound
substructures within the FoF groups are identified
using \textsc{subfind} \citep{Springel2001}. We refer to the most massive subhalo
of a FoF group as ``central'' or ``primary'' and to the remainder as
``satellites''.

APOSTLE galaxies and haloes are identified as bound structures, found
by \textsc{subfind} within $3$ Mpc from the main pair barycentre. We
hereafter refer to the MW and M31 galaxy analogues as
``primaries''. Satellites are identified as galaxies located within
the virial radius of each of the primaries.
The objects of study in this paper have been assigned an identifier in
the form {\tt Vol-FoF-Sub}, where 'vol' {\tt Vol} refers to the corresponding 
APOSTLE volume \citep[ranging from 1 to 12, see Table~2 in][]{Fattahi2016},
and {\tt FoF} and {\tt Sub} correspond to the FoF and
\textsc{subfind} indices, respectively. These indices are computed
for the snapshot corresponding to $z=0$ for LMC analogues (see Tab.~\ref{tab:periapo})
or for the snapshot corresponding to ``identification time'' ($t_{\rm id}$, see Sec.~\ref{SecLMCAssocSats})
for LMC-associated satellites.
We identify  the  stellar mass, $M_*$, of a subhalo with that of all stellar
particles associated  with that system by \textsc{subfind}.

\section{LMC analogues in APOSTLE}
\label{SecLMCanalogs}

Fig.~\ref{fig:LMCimage} illustrates the distribution of dark matter,
gas, and stars in one of the APOSTLE volumes at $z\sim 0$.  The upper
panel illustrates the dark matter distribution, centered at the
midpoint of the ``MW-M31 pair''. The M31 analogue is located in the
upper right part of the panel, whilst the MW analogue is in the bottom
left.  A rectangle shows the area surrounding the MW analogue shown in
the bottom panels, which show the stellar component (left) and gas
(right). The most massive satellite of the MW analogue is situated at
the lower-right in the bottom panels. Note the purely gaseous trailing
stream that accompanies this satellite, invisible in the stellar
component panel. This is one of the ``LMC analogues'' studied in this paper. We focus here on the stellar mass and
kinematics of LMC analogues and their satellites, and defer the study of
the properties of the Magellanic stream-like gaseous features to a
forthcoming paper.


We search for ``LMC analogues'' in APOSTLE by considering first the
most massive satellites closer than 350 kpc to each of the
two primary galaxies in the 12 APOSTLE volumes.  We note that
  this distance is somewhat larger than the virial radius of the
  primaries at $z=0$ ($\sim200$ kpc, see
  Fig.~\ref{fig:LMCorbits}). This prevents us from missing cases of
  loosely-bound LMC analogs that may be past first pericentre at $z=0$
  and just outside the nominal virial boundary of its primary.  This
yields a total of 24 candidates, which we narrow down further by
introducing stellar mass and kinematic criteria, in an attempt to
approximate the present-day configuration of the LMC.

The mass criterion is illustrated in Figure \ref{fig:vmaxmstar}, where
we show the stellar masses of all 24 APOSTLE primaries (circles) and
their corresponding most massive satellites (crosses), as a function
of their maximum circular velocity, $V_{\rm max}$ \citep[see also
figure 7 in][]{Fattahi2016}.  For reference, the stellar mass and
circular velocity of the LMC are marked with a star:
$M_*^{\rm LMC}= 2.5\times10^9$ M$_\odot$
\citep{Kim1998} and $V_{\rm max}^{\rm LMC}=92$ km/s \citep{vanderMarel2014}.
We consider as candidate LMC analogues of each primary the most massive
satellite with
$8.75<\log M_\star/M_\odot<10$; i.e., those in the grey shaded area in
Figure \ref{fig:vmaxmstar}.  This yields a total of $14$ candidates
with maximum circular velocities in the range $55<V_{\rm max}/$km
s$^{-1}<130$. For reference, this velocity range corresponds to a
virial mass range of roughly
$2.5\times10^{10}<M_{200}/M_\odot<4.5\times10^{11}$ for isolated
halos. Of the 14 LMC candidates, we retain only 9 for our analysis
(indicated in red in Figure \ref{fig:vmaxmstar}) after applying an
orbital constraint described in more detail below
(Sec.~\ref{SecLMCOrbits}).

\begin{figure*}
\includegraphics[width=17cm]{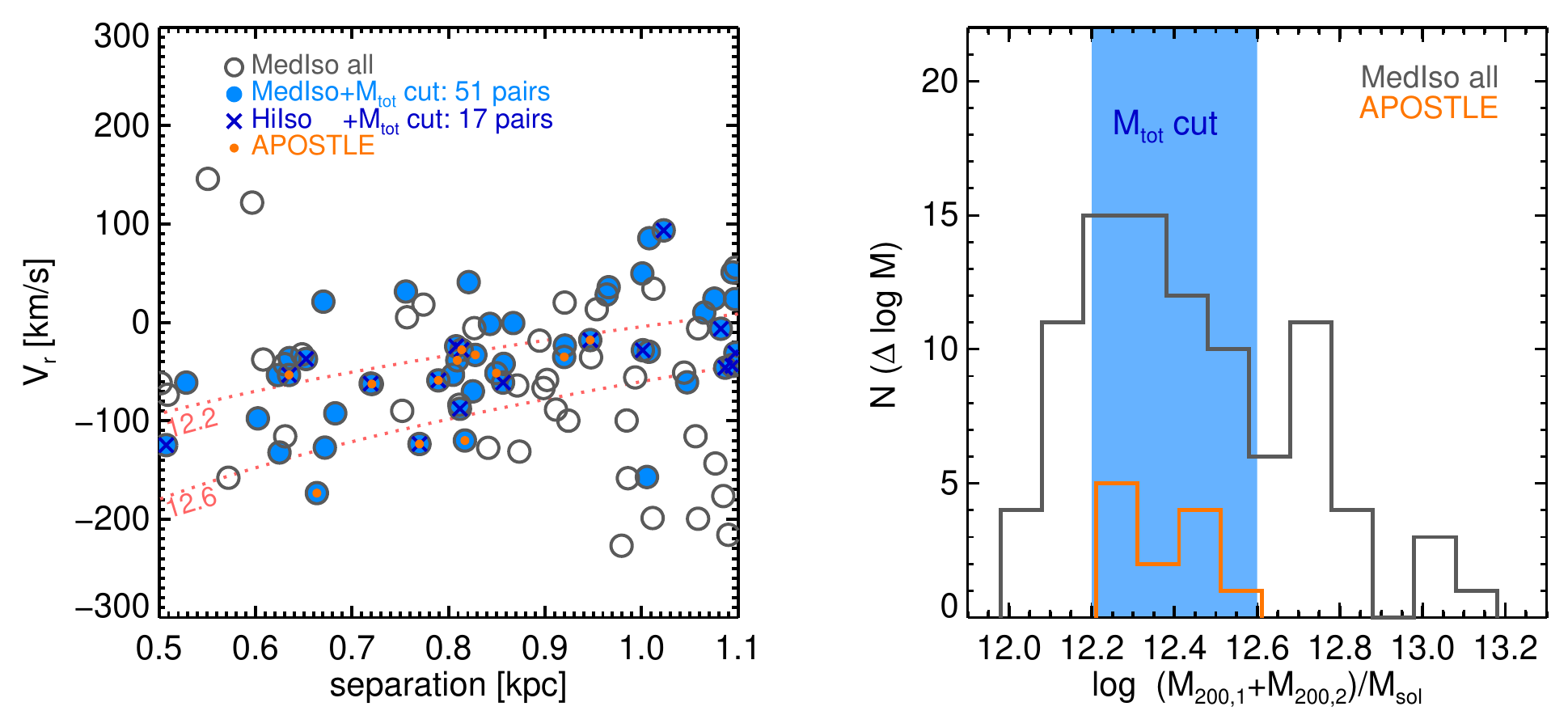}
\caption{\textit{Left}: Separation vs. relative radial velocity of
  halo pair members in DOVE. Open circles indicate MedIso sample galaxies
  (see text for details). Filled blue circles correspond to a
  subsample of MedIso pairs that further satisfies a total mass cut of
  $\log \, ((M_{200,1}+M_{200,2}) /\Msun)=[12.2,12.6]$. Crosses mark
  HiIso sample galaxies with the aforementioned total mass cut. APOSTLE
  pairs, which are a subsample of the MedIso sample, are highlighted
  with small orange circles. Dotted lines indicate timing
  argument solutions for total masses of $\log (M/\Msun)=12.2$ and
  $12.6$, as labelled. \textit{Right}: Total mass,
  i.e. $M_{200,1}+M_{200,2}$, distribution of all the MedIso pairs
  shown in the left hand panel. The shaded blue region indicates the
  additional total mass constraint of
  $\log (M_{\rm tot}/\Msun)=[12.2,12.6]$. An orange histogram shows
  the total mass distribution of APOSTLE pairs.}
    \label{fig:pairs}
\end{figure*}

\begin{figure}
\hspace{-0.2cm}
\includegraphics[width=\columnwidth]{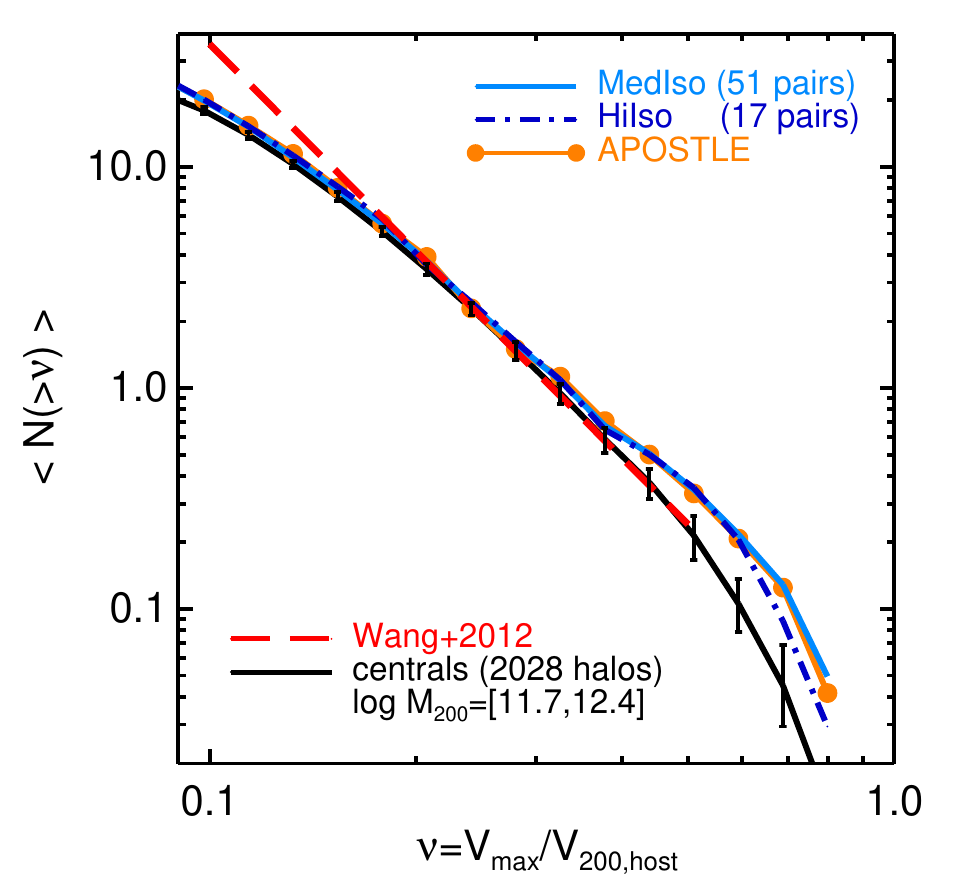}
\caption{Subhalo $\vmax$ function, normalized by the host virial
  velocity $V_{200}$ (i.e. $\nu=\vmax/V_{200,\rm host}$), for
  subhaloes within $r_{200}$ of MW-mass haloes in DOVE. The black line
  corresponds to the average result for 2028 subhaloes around isolated
  haloes with mass $\log(M_{200}/\Msun)=[11.7,12.4]$. The fit to the
  normalized $\vmax$ function from \citet{Wang2012} is shown with the
  red dashed line. The average relation for haloes in the MedIso and
  HiIso pair samples are presented with the light-blue solid line and
  dark blue dashed-dotted line, respectively. The average result for
  subhaloes around APOSTLE primaries is shown with the orange
  connected circles.
  Error bars on the black line indicate the $\pm1\sigma$ dispersion
  around the mean, calculated from 1000 102-halo samples randomly
  drawn from the DOVE catalogue (same number as MedIso primaries).}
    \label{fig:vmaxFnc}
\end{figure}

\subsection{Frequency of LMC-mass satellites}

Fig.~\ref{fig:vmaxmstar} shows that, out of 24 APOSTLE primaries, 
$14$ host 
nearby
satellites massive enough to be comparable to the LMC. 
Of these, 11 are within the virial radius of their host at $z=0$.
This is
a relatively high frequency somewhat unexpected compared with earlier
findings from large cosmological simulations.  Indeed, in the
Millenium-II (MS-II) DM-only simulation only $8$ to $27\%$ of MW-mass
haloes with virial masses between $1$ and $2.5\times 10^{12}\, M_\odot$ are
found to host a subhalo at least as massive as that of the LMC within
their virial radii \citep{Boylan-Kolchin2010}.

This apparent tension motivates us to consider potential environmental
effects that may affect the presence of massive satellites. The Local
Group environment, after all, is characterized by a very particular
configuration, with a close pair of halos of comparable mass
approaching each other for the first time. Could this environment
favor the presence and/or late accretion of massive satellites into
the primaries, compared with isolated halos of similar mass?

We explore this using the DOVE simulation, where we identify pairs of
haloes according to well-defined mass, separation, and isolation
criteria in an attempt to approximate the properties of the Local
Group environment.  We start by selecting haloes with virial masses
$M_{200}>5\times 10^{11}\Msun$ and select those that are within
(0.5-1.1) Mpc of another halo in the same mass range. We impose then a
mass ratio cut of $M_{200,2}/M_{200,1}>0.3$, in order
to retain pairs with comparable mass members, and similar to the
MW-M31 pair. (Here $M_{200,1}$ refers to the virial mass of the more
massive halo of the pair; $M_{200,2}$ to the other.)

We apply next an isolation criterion such that there is no halo (or
subhalo) more massive than $M_{200,2}$ within
$r_{\rm iso}=2.5$ Mpc, measured from the midpoint of the pair. A stricter
isolation criteria is defined by increasing the isolation radius to
$r_{\rm iso}=5$ Mpc. Following \citet{Fattahi2016}, we refer to the
first isolation as ``MedIso'' and to the stricter one as ``HiIso''.

We do not distinguish between centrals and non-centrals in our pair
selection. In fact, in some cases, pair members share the same FoF
group. These are always the two most massive subhaloes of their FoF
group. Our isolation criterion discards pairs of haloes that are
satellites of a more massive halo.

The relative radial velocity vs. separation of all MedIso pairs is
presented in the left panel of Fig.~\ref{fig:pairs} with open
circles. The total mass, $M_{\rm tot}=M_{200,1}+M_{200,2}$, of these
pairs span a wide range as shown by the grey histrogram in
the right-hand panel of Fig.~\ref{fig:pairs}. We further select only
pairs with total mass in the range
$\log (M_{\rm tot}/\Msun)=[12.2,12.6]$, as marked by the blue shaded
region in the right panel. This range includes the total masses of
all APOSTLE pairs (yellow histogram in the right panel). MedIso pairs
that satisfy this total mass criterion are highlighted
with blue filled circles in the left panel. 
This mass cut excludes pairs with the largest total 
masses and most extreme relative radial velocities, which are outliers from the
timing-argument predictions for two point-masses
on a radial orbit approaching each other for the first time (red
dotted curves labelled by the value of $\log M_{\rm tot}/M_\odot$) .

We shall hereafter refer as ``MedIso sample'' to the final sample of
DOVE pairs (with 51 pairs) that satisfy all the above ``Local Group criteria'',
summarized below:
\begin{itemize}
    \item separation: $0.5$-$1.1\Mpc$    
    \item minimum mass of individual haloes: $M_{200}>5\times 10^{11} \Msun$
    \item comparable mass pair members: $M_{200,2}/M_{200,1}>0.3$
    \item total mass of pairs: $\log
      (M_{200,1}+M_{200,2})/M_\odot=[12.2,12.6]$
    \item MedIso isolation: $r_{\rm iso}=2.5 \Mpc$
\end{itemize}
The final ``HiIso sample'', with 17 pairs, satisfies all the above
conditions but has a stricter isolation criterion of $r_{\rm
  iso}=5 \Mpc$. These are marked with crosses in  Fig.~\ref{fig:pairs}.

APOSTLE pairs are a subsample of the MedIso group, but with extra
constraints on the
relative radial and tangential velocity between the primaries,
as well as on the Hubble flow velocities of objects surrounding the primaries out to 4 Mpc
\citep[see][for details]{Fattahi2016}. They are marked with small orange filled
circles in the left panel of Fig.~\ref{fig:pairs}, and their total
mass distribution is shown by the orange histogram in the right-hand
panel of the same figure.

We compare in Fig.~\ref{fig:vmaxFnc} the abundance of (massive)
subhaloes around APOSTLE primaries with those of MedIso and HiIso
pairs, as well as with all isolated MW-mass haloes in DOVE. The latter
is a ``control sample'' that includes all central subhalos with
$11.7<\log (M_{200}/M_\odot)<12.4$ found in the DOVE cosmological
box. This mass range covers the range of masses of individual pair
members in APOSTLE and in the MedIso sample.

Fig.~\ref{fig:vmaxFnc} shows the scaled subhalo $\vmax$ function,
i.e. $N (>\nu) \equiv N(>\vmax/V_{\rm 200,host} )$, averaged over host
haloes in various samples. We include all subhaloes within $r_{200}$
of the hosts. The scaled subhalo $\vmax$ function
of the control sample (solid black curve) is consistent with the fit
from \citet{Wang2012}, who used a number of large cosmological simulations and
a wide halo mass range (red dashed curve). The turnover at $\nu < 0.15$
is an artifact of numerical resolution, which limits our ability to
resolve very low mass haloes.

Interestingly, Fig.~\ref{fig:vmaxFnc} shows that, on average, our
various paired samples (MedIso, HiIso, APOSTLE) have an overabundance
of massive subhaloes relative to average isolated
$\sim10^{12}\, M_\odot$ haloes. Indeed, the chance of hosting a
massive subhalo with $\nu>0.6$ almost doubles for halos in LG-like
environments compared with isolated halos.

Error bars on the $\nu$ function of the control sample represent the 
$\pm 1\sigma$ dispersion around the average, computed by randomly drawing 102 halos 
(as the total number of halos in the MedIso paired sample) from the sample of 2028 
DOVE centrals, 1000 times. 
We find that only 2/1000 realizations reach the $\langle N(\nu)\rangle$ 
measured for APOSTLE pairs at $\nu=0.6$, proving the robustness of the result.

We note that the overabundance of massive subhaloes in halo pairs
persists when altering the isolation criterion (HiIso vs. MedIso) or
when using a more restrictive selection criteria
on the relative kinematics of the halos and the surrounding Hubble flow
 (APOSTLE
vs. MedIso). We have additionally checked that imposing tighter
constraints directly on the MedIso sample ($V_{r}=[-250,0]\,\kms$,
$d=[0.6,1]\,\Mpc$) does not alter these conclusions.
Moreover, we have explicitly checked that the higher frequency of massive satellites 
found in the  paired halo samples is not enhanced by the most massive primaries in 
the host mass range considered ($11.7<\log (M_{200}/M_\odot)<12.4$).
Therefore, the main environmental driver for the overabundance 
of massive subhalos in Local Group-like environments seems to be the 
presence of the halo pair itself.

This result is consistent with that of \citet{Garrison-Kimmel2014},
who report a global overabundance of subhalos in Local Group-like
pairs compared to isolated MW-like halos.  However, we caution that
some of the volumes analysed by these authors were specifically
selected to contain LMC-like objects, so it is not straightforward to
compare our results quantitatively with theirs.  We conclude that
haloes in pairs such as those in the Local Group have a genuine
overabundance of massive satellites compared to isolated
halos. LMC-like satellites are thus not a rare occurrence around Milky
Way-like hosts.

\subsection{The orbits of LMC analogues}
\label{SecLMCOrbits}

\begin{figure}
\centering
\includegraphics[width=\linewidth]{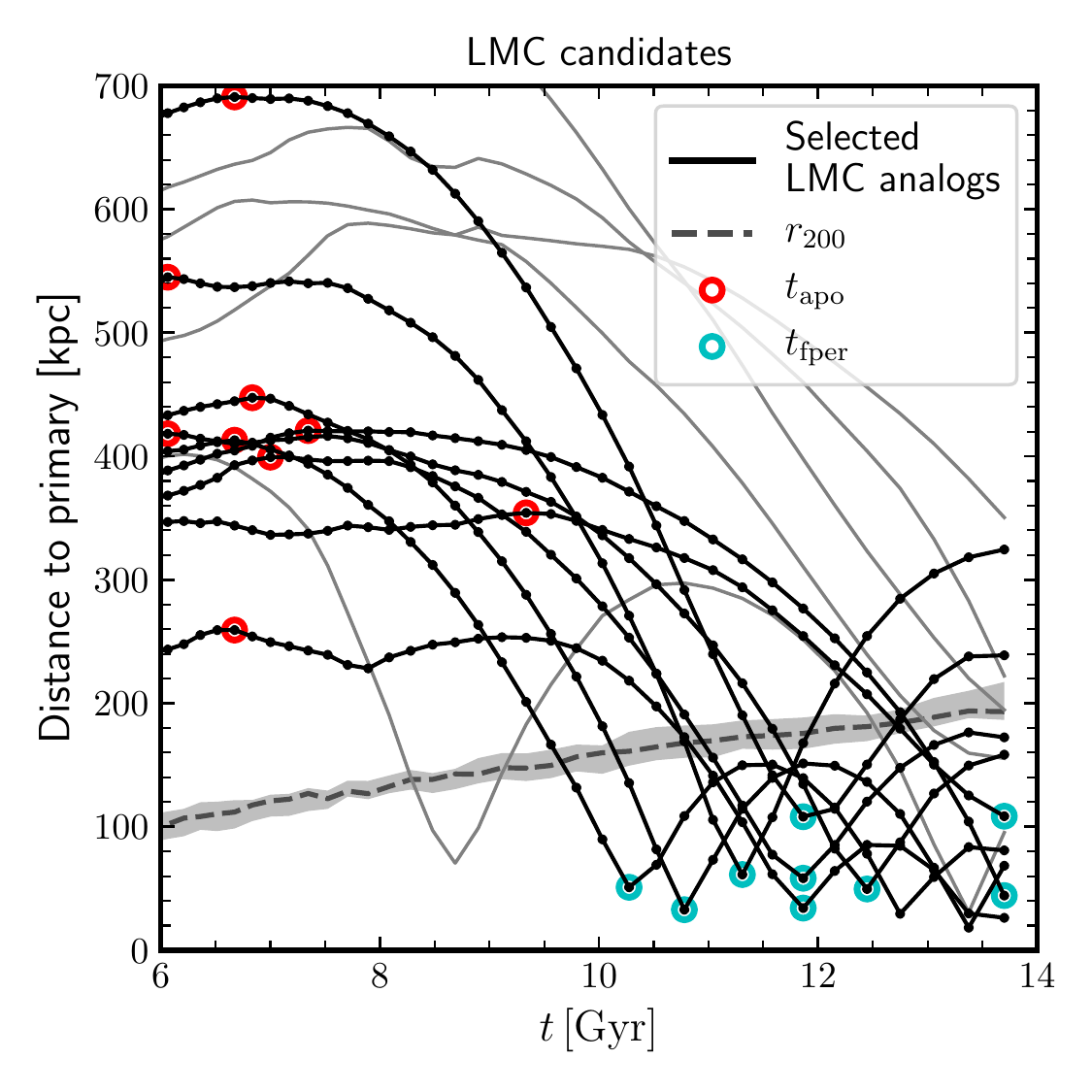}
\caption{Radial distance to the primary versus time, for the $14$ LMC
  analogues identified in Fig.~\ref{fig:vmaxmstar}. The final $9$ LMC
  analogues analyzed in this work are shown in black, while the rest of
  candidates are shown in gray. A cyan circle highlights the time when
  the LMC analogue is at first pericentre. Red circles mark the time of
  ``turnaround'' (first apocentre).  The average time evolution of the
  virial radius of the primaries is shown with a dashed line (median
  and 25-75 percentiles).}
 \label{fig:LMCorbits}
\end{figure}

LMC analogues should not only match approximately the LMC's
stellar mass (Fig.~\ref{fig:vmaxmstar}) but also its orbital
properties and dynamical configuration. We therefore refine our identifying
criteria by inspecting the orbits of the $14$ LMC-analogue candidates,
shown in Fig.~\ref{fig:LMCorbits}. We shall retain as LMC analogues only
candidates that have been accreted relatively recently (i.e., those
that undergo the first pericentric passage at times
$t_{\rm fper} > 10$ Gyr, or $z_{\rm fper}<0.37$) and that, in
addition, have pericentric distances $r_{\rm peri} \lesssim 110$ kpc.

Fig.~\ref{fig:LMCorbits} shows that $9$ out of the $14$ original
candidates satisfy these conditions (this final sample of LMC analogues
is shown in red in Fig.~\ref{fig:vmaxmstar}).  
We highlight the orbits of the selected candidates in
Fig.~\ref{fig:LMCorbits} using black curves, where the cyan and red
 circles indicate their pericentres and apocentres,
respectively\footnote{These apocentres are actually best understood as
  ``turnaround radii''; i.e., as the maximum physical distance to the
  primary before infall.}.  The rest of the candidates that do not meet
the orbital criteria are shown in grey. Of these, we find only one
case with a very early first pericentre (at $t\sim 8.7$ Gyr) that is
at present on its second approach. The others have either not yet
reached pericentre by $z=0$ or have very large ($\sim 200$ kpc)
pericentric distances. The APOSTLE LMC analogues are thus recently
accreted satellites, in line with the conclusions of
\citet{BoylanKolchin2011a}, who find that $50\%$ of massive satellites
in the MS-II DMO simulation have infall times in the last 4 Gyrs.

We list the individual pericentric and apocentric distances of each of
our 9 LMC analogues in Table~\ref{tab:periapo}. The median
pericentre is $\sim 60$ kpc, in good agreement with the pericentre
estimates for the LMC at $\sim 50$ kpc. The analogues show a wide range
of apocentres, which extend from $\sim 260$ kpc all the way to
$700$ kpc, with a median of $\sim 420$ kpc. The typical orbit of LMC
analogues in our sample is therefore quite eccentric, with a median
eccentricity $\epsilon \equiv r_{\rm peri}/r_{\rm apo} = 0.12$.

One may use these typical values to draw inferences regarding the past
orbital history of the LMC around the MW. For example, taking the
LMC's current Galactocentric radial distance as pericentre distance
(i.e., $r_{\rm peri}^{\rm LMC} = 49.9$ kpc; see
Table~\ref{tab:lmcdata}) the median eccentricity, $\epsilon=0.12$,
suggests an apocentre for the LMC of $r_{\rm apo}^{\rm LMC} \sim 408$
kpc before starting its infall towards our Galaxy.

The large apocentric distances discussed above allow the $9$ LMC analogues
to acquire substantial angular momentum through tidal torqing by the
nearby mass distribution. Table~\ref{tab:periapo} lists the specific orbital angular
momentum of each simulated LMC analogue at first pericentre normalized
by the virial value ($r_{200}\times V_{200}$) of the corresponding
primaries measured at the same time.  The median of the sample is
$|\vec{l}_{\rm orb}|/(r_{200} \times V_{200}) = 0.64$, in good agreement with
the value ($\sim 0.54$) estimated assuming the latest LMC kinematics
constraints from Table~\ref{tab:lmcdata} \citep{Kallivayalil2013} and
a virial mass $M_{200} = 1 \times 10^{12}\; \rm M_\odot$ for the
MW. Under the condition of recent infall, the large orbital spin of
the LMC around the Galaxy is not difficult to reproduce within $\Lambda$CDM
\citep[see also ][]{BoylanKolchin2011a}.

\begin{figure}
  \centering
\includegraphics[width=\linewidth]{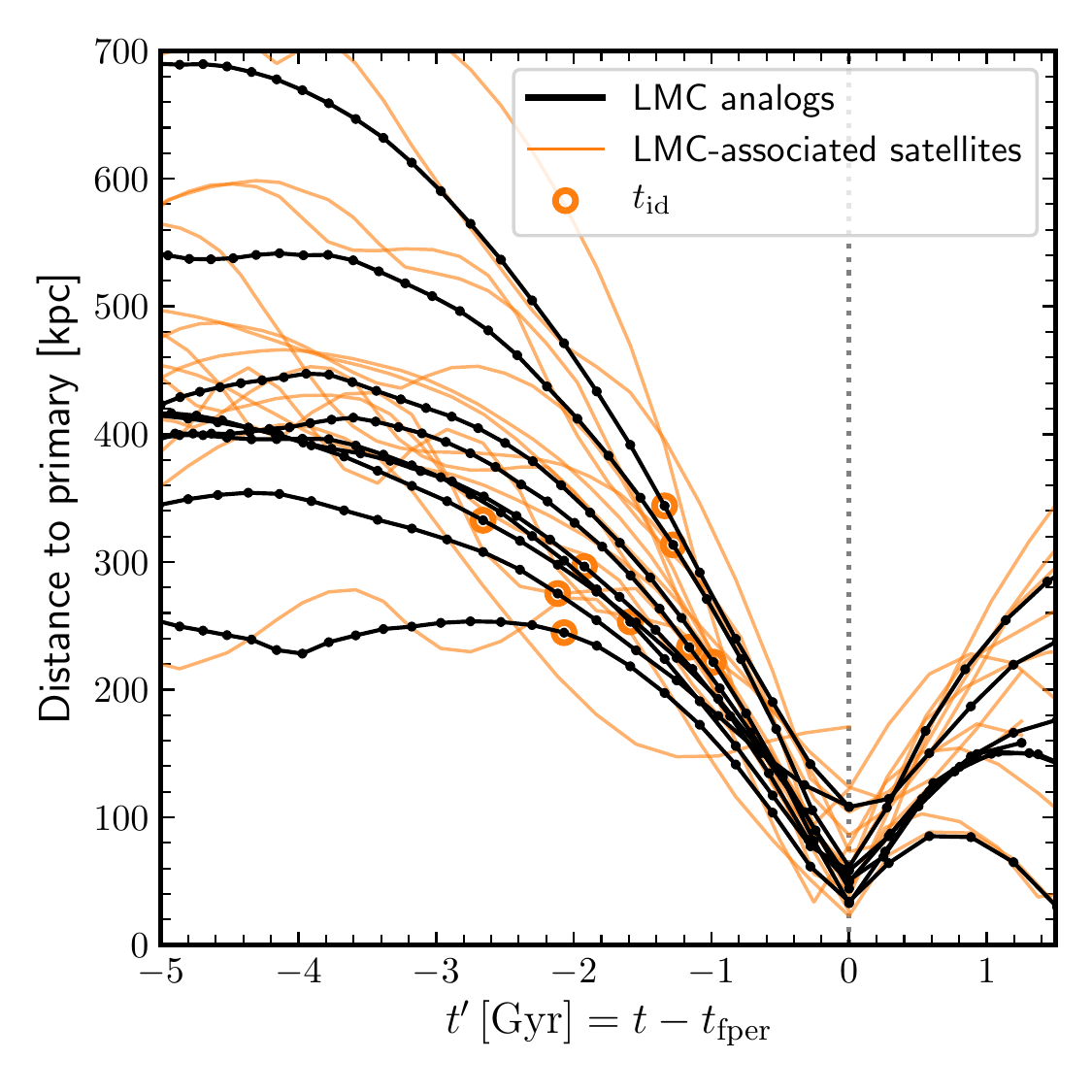}
\caption{Radial distance to the primary versus time for LMC analogues
  (black) and LMC-associated satellites (orange). The time axis has
  been shifted so that all objects are at their first pericentre at
  $t'=0$. The times at which LMC-associated satellites have been
  identified around their corresponding LMC analogues ('identification
  time', $t_{\rm id}$), are highlighted with orange circles.  }
\label{fig:satorbits}
\end{figure}

\begin{table*}
\centering
\caption{Observational data assumed in this work for the LMC. Stellar
  mass, Galactocentric position and velocity, Galactocentric radial
  distance and magnitude of the specific orbital angular momentum
  vector.  Galactocentric Cartesian position has been computed from
  the RA, dec and $(m-M)$ values quoted in the latest data being made
  available by the \citet{McConnachie2012} compilation. Galactocentric
  velocities have been computed assuming a heliocentric line-of-sight velocity of
  $V_{\rm los}= 262.3$ km/s \citep{vanderMarel2002} and proper motions
  $\mu_W=-1.899$ mas/yr, $\mu_N=0.416$ mas/yr
  \citep{Kallivayalil2013}.  We assume a distance of the Sun from the
  Milky Way of $R_\odot=8.29$ kpc, a circular velocity of the local
  standard of rest (LSR) of $V_0=239$ km/s \citep{McMillan2011}, and a
  peculiar velocity of the Sun with respect to the LSR of
  $(U_\odot,V_\odot,W_\odot) = (11.1, 12.24,7.25)$ km/s
  \citep{Schonrich2010}.  }
\begin{tabular}{ l l l l l l l l l }
\toprule
$M_{*}$/M$_\odot$&$X$/kpc  & $Y$/kpc   & $Z$/kpc   & $V_X$/km s$^{-1}$   & $V_Y$/km s$^{-1}$   & $V_Z$/km s$^{-1}$   &   Distance/kpc  &  $|\vec{l_{\rm orb}}|$/(kpc km s$^{-1}$)   \\
\midrule
2.5$\times 10^9$  & -0.58 & -41.77  & -27.47 & -85.41 & -227.49  & 225.29 & 49.99 & 16221.26\\
\bottomrule
\end{tabular}
\label{tab:lmcdata}
\end{table*}


\begin{table}
\centering
\small
\caption{Orbital characteristics of the $9$ LMC analogues presented in this work.
Column 1 indicates the LMC analogue identifier. 
LMC analogues are identified with a label in the form {\tt Vol-FoF-Sub}, that indicates the corresponding 
APOSTLE volume, as well as the FoF and \textsc{subfind} indices of the object in the $z=0$ snapshot.
Column 2  indicates  the redshift at which the LMC analogue's corresponding satellites have been identified (`identification time' $t_{\rm id}$, see Sec.~\ref{SecLMCAssocSats}).
Throughout this paper, LMC analogues and their respective satellites are shown in a same color
consistently in all figures. LMC analogues in this table are ordered by this color, from red to dark blue.
 Subsequent columns indicate  the LMC analogue's pericentric distance,
 apocentric distance, orbital eccentricity ($\epsilon = r_{\rm peri}/r_{\rm apo}$), and magnitude of the specific orbital angular momentum vector $\vec{l}_{\rm orb}$ normalized by ($r_{200}\times V_{200}$) of its corresponding primary.
}
\begin{tabular}{ l l l l l l }
\toprule
Label & $z_{\rm id}$ & $r_{\rm peri}$/kpc & $r_{\rm apo}$/kpc & $\epsilon$ & $|\vec{l_{\rm orb}}|$/($r_{200}\times V_{200}$)   \\
\midrule
\midrule
5-2-2   &0.503   &    51.00    &    412.94    &    0.12    &   0.72 \\
2-1-3   &0.399  &    32.83    &    447.37    &    0.07    &    0.51 \\
1-1-1   &0.366  &    61.29    &    544.97    &    0.11    &    0.64 \\
12-1-4   &0.399 &    34.14    &    259.27    &    0.13    &  0.33    \\
11-1-4   &0.333  &    58.32    &    399.28    &    0.15    &  0.66  \\
11-1-3   &0.302 &    49.59    &    418.20    &    0.12    &  0.51    \\
10-1-2   &0.241  &    44.25    &    420.76    &    0.11    &  0.28     \\
1-2-2    &0.183  &    108.52    &    354.17    &    0.31    &  0.64 \\
3-1-1    &0.302 &    108.12    &    690.90    &    0.16    &    0.77  \\
\midrule
Median & &  50.99 &  418.19  & 0.12  & 0.64  \\
\bottomrule
\end{tabular}
\label{tab:periapo}
\end{table}

\section{LMC-associated satellites in APOSTLE}
\label{SecLMCAssocSats}
 
Given the relatively high masses of the LMC analogues, we expect them to
harbour their own population of satellite dwarfs. We identify them in
the simulations as follows.  We first trace their orbits back from
pericentre until they are $\sim 100$ kpc away from the virial boundary
of the primary. At that time in the orbit, referred to as
``identification time", or $t_{\rm id}$, we flag as ``LMC satellites''
all luminous subhalos within $100$ kpc of each LMC analogue.
We include all luminous subhalos; i.e., with at least 1 star
  particle, unless otherwise specified.

The procedure yields a combined total of $16$ satellites for the $9$
LMC analogues.  Only one LMC analogue is ``luminous satellite-free''
at $t_{\rm id}$.  We have traced the orbital evolution of the
  LMC satellites in time and have confirmed that all are bound to
  their LMC analogues, at least until first pericentre.  One of the
satellites merges with its LMC analogue before the latter reaches
first pericentre. Our final sample therefore consists of $15$
LMC-associated satellites.

Using merger trees, we trace back and forth in time
each of the LMC-associated satellites. We show their orbits in
Fig.~\ref{fig:satorbits} with orange curves, together with those of
their respective primaries. Times in this figure have been shifted so
that $t'=t-t_{\rm fper}=0$ corresponds to that of the snapshot
corresponding to the closest approach of each LMC analogue.
``Identification times" for each LMC analogue are highlighted with orange
circles in Fig.~\ref{fig:satorbits}.

This figure shows that, at first pericentre, LMC-associated satellites
remain very close in radial distance to their corresponding LMC
analogue, although they may evolve differently afterwards. This implies,
as suggested in Sec.~\ref{SecIntro}, that any MW satellite associated with
the LMC should be found at a close distance from the LMC today. We shall
return to this issue in Sec.~\ref{SecLMCIdCrit}.

Hereafter, all the results shown correspond to $t_{\rm fper}$, unless otherwise stated.

\subsection{Projected position and orbital angular momentum}
\label{ssec:phasespace}

\begin{figure*}
\includegraphics[width=\linewidth]{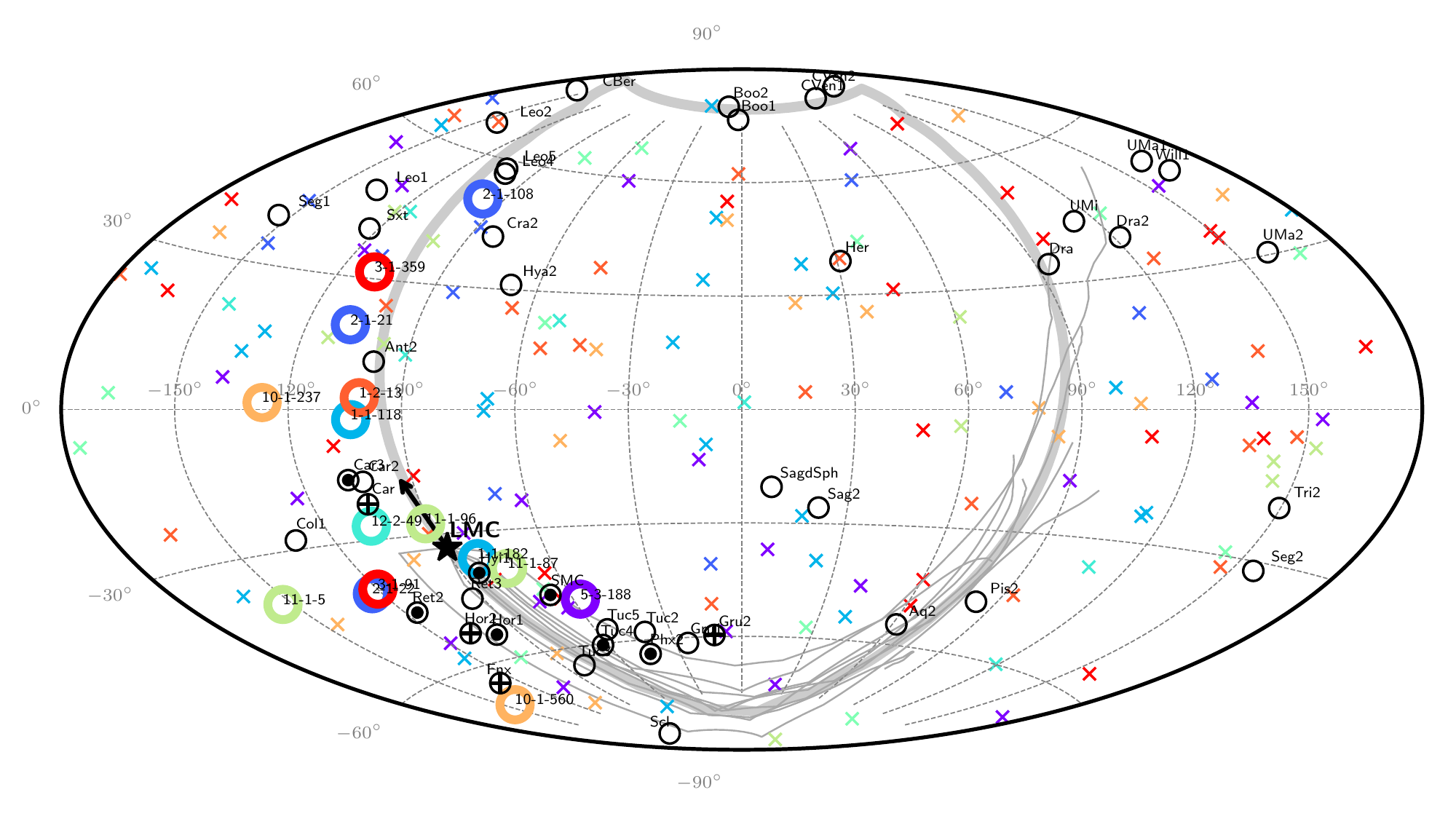}
\includegraphics[width=\linewidth]{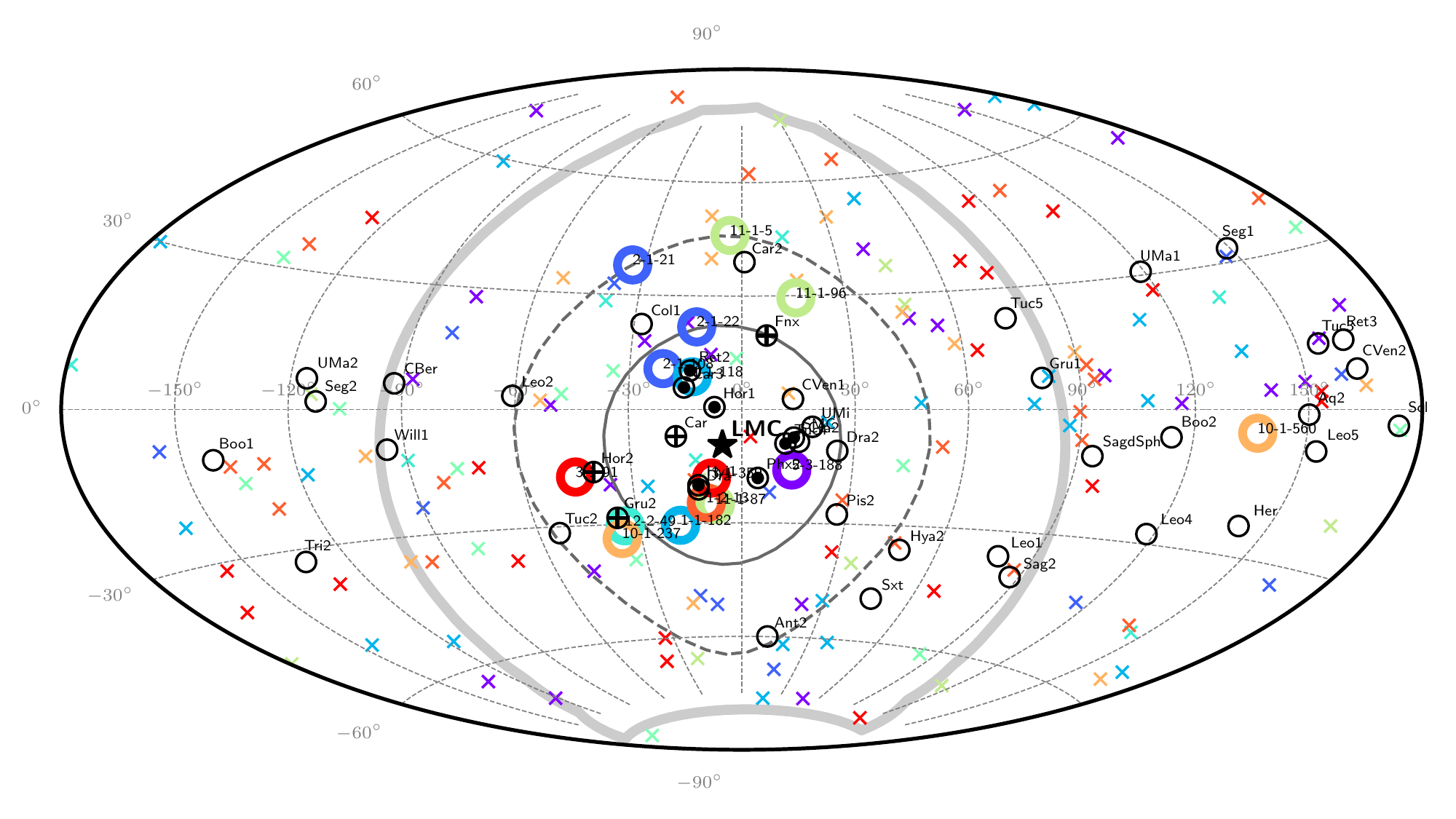}
\caption{Position (top) and orbital angular momentum direction
  (bottom) of satellites of the primary haloes relative to the LMC
  (black star), in Galactocentric coordinates.  LMC-associated
  satellites are shown as large open circles with labels. The rest of
  satellites of the primary are shown as crosses. Satellites belonging
  to the same primary are shown in the same color.  Coordinates
  systems are rotated such that the positions and orbital poles of LMC
  analogues coincide with the corresponding observed values for the LMC,
  indicated with a large star.  Observed MW satellites are shown as
  open black circles with labels. MW satellites highlighted with a
  filled circle or a cross are those deemed likely LMC associates
  according to the discussion in Sec.~\ref{SecLMCIdCrit}.  Thin gray
  lines in the top panel show the individual orbital trajectories of
  each of the 9 LMC analogues. An arrow indicates the direction of
  motion of the LMC along the trajectory.  In the bottom panels, for
  reference, we show circles centred on the LMC with aperture
  $32^\circ$ and $55^\circ$, respectively (see text for details).  }
    \label{fig:posaitoff}
\end{figure*}

\begin{figure*}
\includegraphics[width=0.55\linewidth]{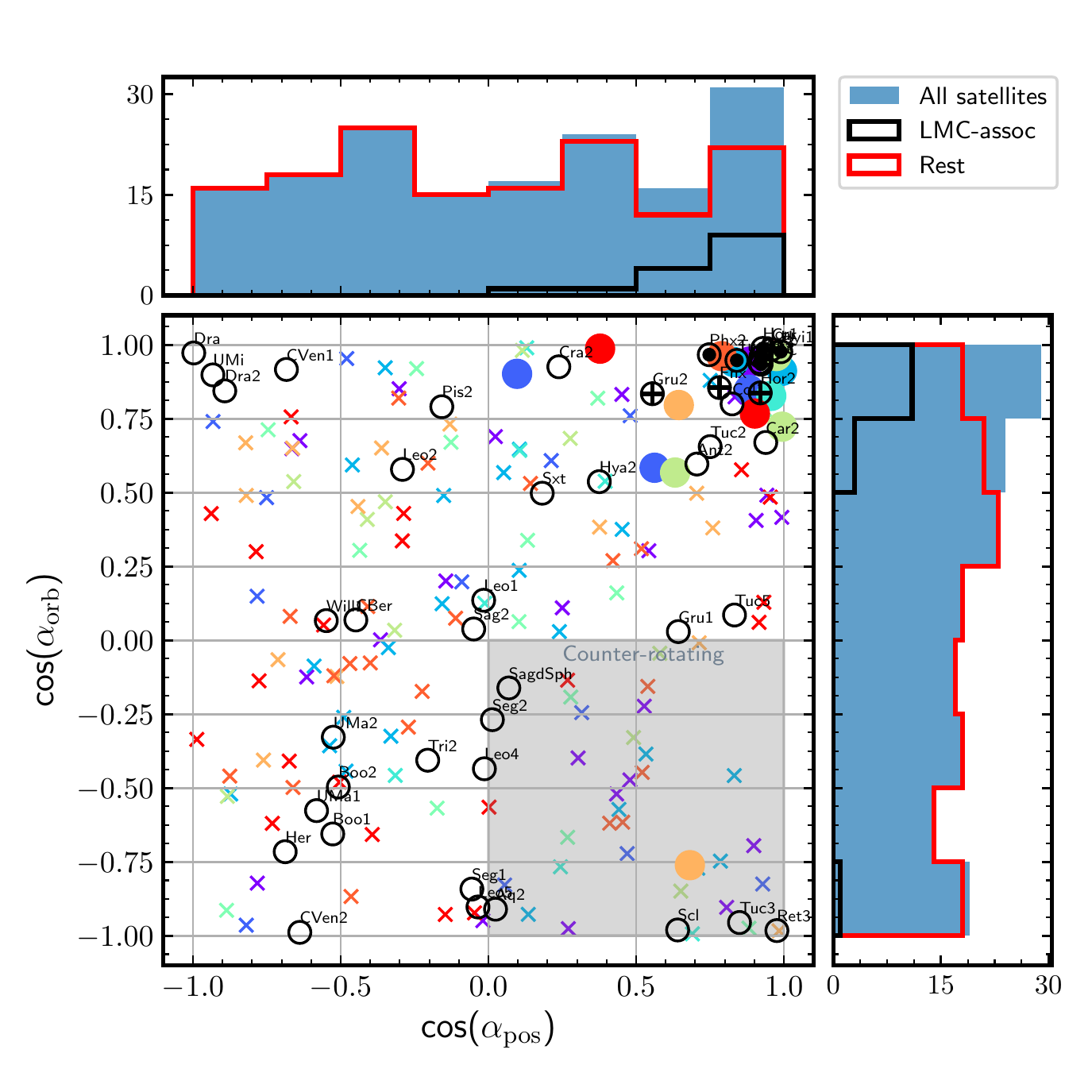}
\includegraphics[width=0.44\linewidth]{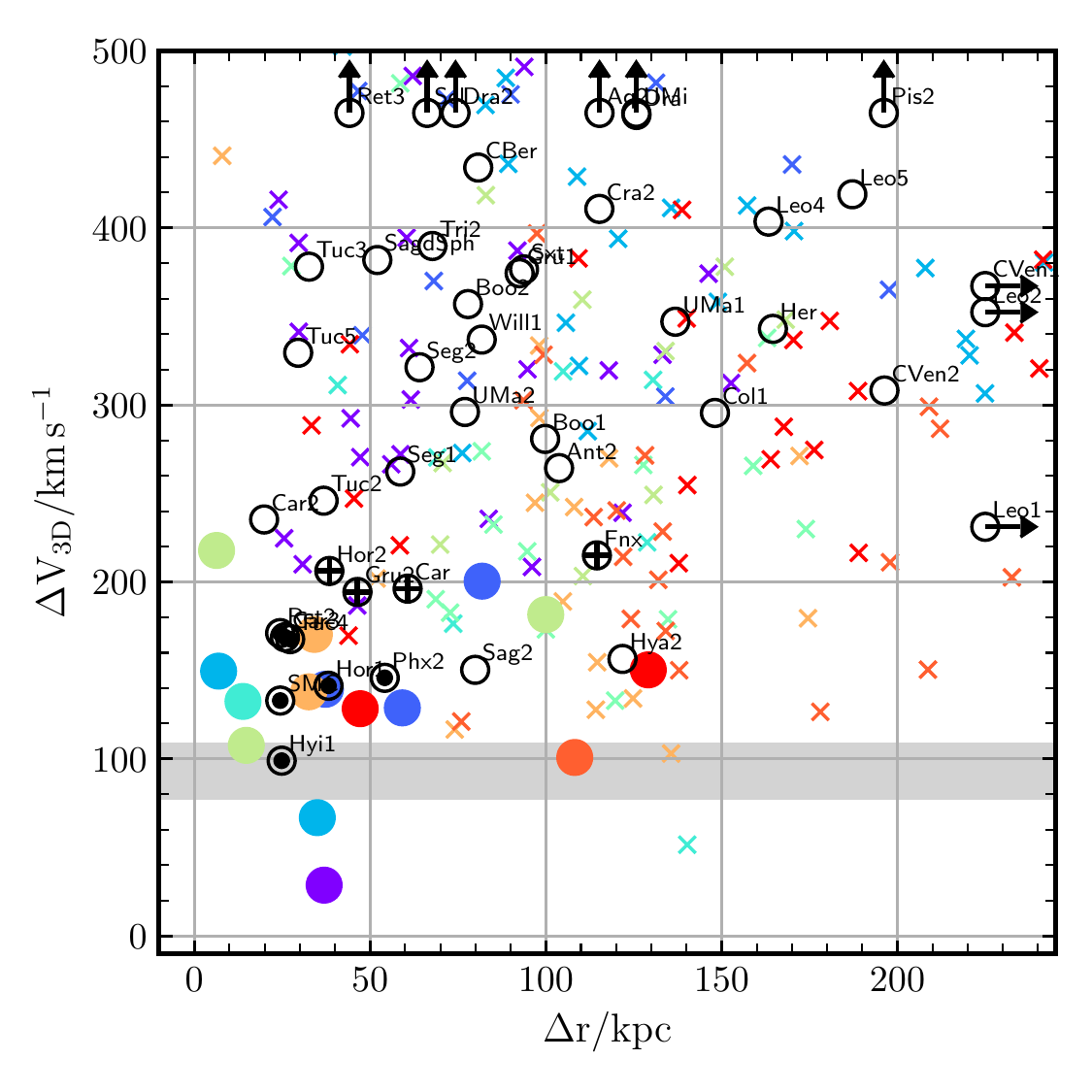}
\caption{ \textit{Left:} Angular separation between the position
  vector of satellites and that of the LMC analogue, versus angular
  separation between the orbital angular momentum direction of
  satellites and that of the LMC analogue.  LMC-associated satellites
  from APOSTLE are shown as colored filled circles, and the rest of
  satellites as crosses.  Histograms show the distribution along the
  axes of the different samples of satellites (i.e., all satellites,
  LMC-associated satellites, and the rest of satellites of the
  primary).  \textit{Right:} Radial distance and 3D velocity of
  LMC-associated satellites relative to that of their LMC analogue, at
  first pericentre. 
  A shaded band indicates the 25-75 percentile range of $V_{\rm max}$ 
  values for LMC analogues, as a reference. 
  Color-coding in both panels is the same as in
  Fig.~\ref{fig:posaitoff}.  For comparison, MW satellites are shown
  as black open circles with labels.  MW satellites highlighted
  with a filled circle or a cross are those deemed likely LMC
  associates according to the discussion in Sec.~\ref{SecLMCIdCrit}.  }
    \label{fig:cosalfa}
\end{figure*}
The top panel of Figure \ref{fig:posaitoff} shows an Aitoff projection
of the sky position of all satellites associated with the primaries
hosting LMC analogues at the time of first pericentre. Each of the
coordinate systems of the 9 LMC analogues has been rotated so that the
LMC analogue is at the same Galactocentric position in the sky as the
observed LMC and the orbital angular momentum vector of the LMC analogue
is parallel to that of the observed LMC (see Table~\ref{tab:lmcdata}
for the position and velocity data assumed for the LMC). The position
of the LMC (analogue and observed) is marked with a star, while
LMC-associated satellites are shown as large colored open circles with
labels. The remainder of the satellites of each primary are shown as
colored crosses. A different color is used for each of the 9 primaries
containing LMC analogues.

For comparison, observed MW satellites\footnote{We show data for all
  known MW satellites within 300 kpc with measured kinematic data, including a few cases where it
  is unclear if the system is a dwarf galaxy or a globular cluster
  \citep[see][]{McConnachie2012}. See Table~\ref{TabScores} 
  for a listing of the objects considered and the corresponding data references.}  are overplotted as small black open
circles with identifying labels. In addition, a thick gray line marks
the LMC's orbital plane and an arrow indicates the direction of motion
along this line.  Individual thin gray lines show each of the LMC
analogues' orbital paths, starting at ``turnaround'' (apocentre) and
ending at pericentre.  One interesting result is that APOSTLE LMC
analogues mostly follow the same orbital plane during their infall onto
the primary.  This is in good agreement with \citet{Patel2017b}, who find
LMC-mass satellites in the Illustris simulations with late accretion
times generally conserve their orbital angular momentum up to $z=0$.

The spatial distribution in the sky of the LMC-associated satellites
clearly delineates the orbital plane of the LMC, which appear to
spread more or less evenly along the leading and trailing section of
the orbital path, as expected if LMC satellites were to accompany the
orbit of the LMC.  The bottom panel of Figure \ref{fig:posaitoff}
shows that this is indeed the case: the instantaneous direction of the orbital
angular momentum vectors (or orbital 'poles') of LMC-asociated
satellites at $t_{\rm fper}$ seems to coincide rather well with that of the LMC
itself. Again the coordinate system of each LMC analogue has been
rotated\footnote{Here longitude coordinates have been rotated by 180
  degrees to show the angular momentum of the LMC at the centre of the
  Aitoff diagram.} such that the LMC analogue's orbital pole aligns with
that of the observed LMC, marked with a star.


\begin{figure*}
\includegraphics[width=\linewidth]{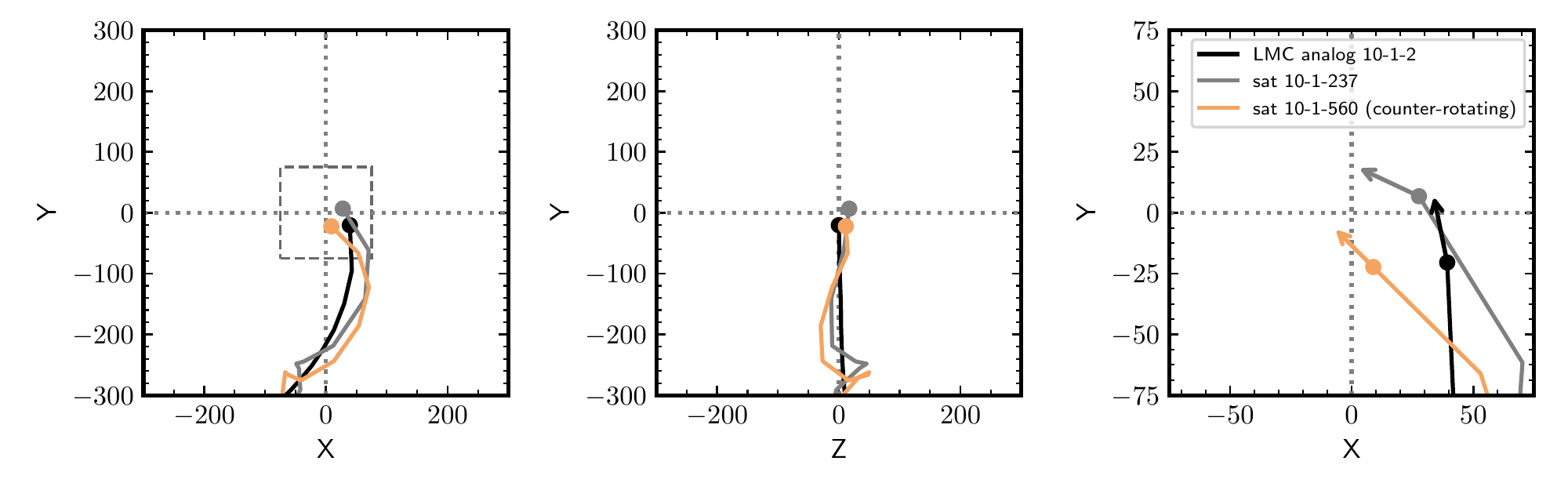}
\caption{Orbital trajectory of the LMC-associated satellite (labelled
  10-1-560, see Fig.~\ref{fig:posaitoff}) that appears to be
  counter-orbiting with respect to its LMC analogue at the time of first
  pericentre (in orange).  The trajectory of the LMC analogue is shown
  in black. A second satellite of the same LMC analogue is shown in
  grey.  The reference system is centred on the primary galaxy, and
  the orbital plane of the LMC analogue is chosen as the XY plane.  The
  rightmost panel is a zoomed-in view of the region enclosed in a
  rectangle in the leftmost panel. Arrows in the right-most panel
  indicate the direction of the instantaneous velocity vectors of each
  satellite at the final time.  }
\label{fig:counterrot}
\end{figure*}

The clustering of the orbital poles of LMC-associated satellites is to
be expected although it is perhaps less tight than assumed in earlier
work \citep[see Fig.5 of][]{Kallivayalil2018}.
Indeed, some satellites are found to have orbital poles that differ
from that of the LMC by as much as $\sim55$ degrees (shown as a
dashed-line circle for reference), with a median value of $\sim 32$
degrees (shown as a solid-line circle).

The spatial and pole distributions on the sky of LMC-associated
satellites in APOSTLE are consistent with the location of the bulk of
the debris from the cosmological dark matter-only LMC analogue studied
first in \citet{Sales2011,Sales2017} and compared to \textit{Gaia} data in
\citet{Kallivayalil2018}. However, we find also a surprising result
here: there is the case of a simulated satellite whose orbital pole is
nearly 180 degrees away from its LMC analogue's. In other words, this
satellite appears to be ``counter-rotating'' the Milky Way relative to
the LMC (see orange open circle labelled 10-1-560 in
Fig.~\ref{fig:posaitoff}). We shall explore this case in more detail
in Sec.~\ref{ssec:counterrot}.

One conclusion from Fig.~\ref{fig:posaitoff} is that the orbital
pole condition leaves many MW satellites as potentially associated
with the LMC. It is therefore important to look for corroborating
evidence using additional information, such as positions and
velocities. We explore this in Fig.~\ref{fig:cosalfa}, where the left
panel shows the cosine of the angle between different directions that
relate the LMC with its satellites. The x-axis corresponds to the
angular distance ($\alpha_{\rm pos}$) between the position of the LMC
analogue and other satellites; the y-axis indicates the angular distance
($\alpha_{\rm orb}$) between their corresponding orbital poles.

Satellites associated with LMC analogues are shown with colored circles
in Fig.~\ref{fig:cosalfa}, and are compared with those of MW satellites
with available data (open black circles). The former are clearly quite
close to the LMC both on the sky in position (most have
$\cos \alpha_{\rm pos}>0.5$), and also have closely aligned orbital
poles (most have $\cos \alpha_{\rm orb}>0.5$).

What about the other satellites, which were not associated with the
LMC analogues before infall? Are their positions and/or kinematics
affected by the LMC analogue? Apparently not, as shown by the small
colored crosses in Fig.~\ref{fig:cosalfa} and by the histograms at
the top and right of the left-hand panel of the same figure. Filled
blue histograms show the distribution of each quantity (for simulated
satellites) on each axis. These show a small enhancement towards small
values of $\alpha_{\rm pos}$ and $\alpha_{\rm orb}$, but the
enhancement is entirely due to the satellites associated with the LMC
analogues (black histograms). Subtracting them from the total leaves the
red histogram, which is consistent with a flat, uniform
distribution. In other words, neither the angular positions nor the
orbital angular momentum directions of non-associated satellites seems
to be noticeably affected by a recently accreted LMC analogue.

Besides the projected distance and orbital pole separation shown on
the left panel of Fig.~\ref{fig:cosalfa}, our results also indicate
that satellites associated with the LMC analogues remain close in
relative distance and velocity (something already hinted at when
discussing Fig.~\ref{fig:satorbits}). This is shown in the right-hand
panel of Fig.~\ref{fig:cosalfa}, where we plot the relative velocity
($\Delta V_{\rm 3D}$)
and distance  ($\Delta r$) between all satellites of the primary and the LMC
analogue. Satellites associated with the analogues (filled circles)
clearly cluster towards small $\Delta r$ and small $\Delta V_{\rm 3D}$, with a
median $\Delta r$ of just $\sim 37$ kpc and a median $\Delta V_{\rm 3D}$ of
just $\sim 138$ km/s. We shall use these results to refine our criteria for
identifying LMC-associated satellites in Sec.~\ref{SecLMCIdCrit},
after considering first the peculiar case of a counter-rotating
satellite.

\subsection{A counter-rotating LMC-associated satellite}
\label{ssec:counterrot}

We turn our attention now to the ``counter-rotating'' satellite
highlighted in the Aitoff projection in Fig.~\ref{fig:posaitoff}
(orange open circle labelled 10-1-560), which appears at
$\cos (\alpha_{\rm orb})\sim -0.75$ in the left panel of
Fig.~\ref{fig:cosalfa}. This is clearly an outlier relative to all
other satellites associated with LMC analogues. What mechanism could
explain this odd orbital motion?

With hindsight the explanation is relatively simple, and may be traced
to a case where the amplitude of the motion of a satellite around the
LMC analogue is comparable to the pericentric distance of the latter
around the primary host. This is shown in Fig. \ref{fig:counterrot},
which plots the orbital trajectory of satellite 10-1-560 in a
reference frame centred on the primary and where the XY plane is
defined to coincide with the orbital plane of the LMC analogue. The LMC
analogue is shown in black, and its two satellites in grey and
orange. In all panels, a line shows the trajectory of each object
starting at early times and finalizing at first pericentre (marked
with a circle), which, in this particular case, corresponds with the
last snapshot of the simulation, at $z=0$. The left and middle panels
show the XY and ZY projections of the trajectories in a box $600$ kpc
on a side. The right-hand panel shows a zoomed-in XY view $150$ kpc on
a side, where the arrows indicate the projections of the instantaneous
velocity vectors at first pericentre.

The velocity vectors explain clearly why satellite 10-1-560 appears to
counter-rotate: when the relative ``size'' of the LMC satellite system
is comparable to the pericentric distance of the LMC orbit, the orbital
motion may appear to carry an LMC satellite on an instantaneous orbit
that shares the same orbital plane but that goes around the primary
centre on the opposite side. We find this instance in only one out of
the $15$ satellites we identified and tracked.  This is thus a
possible but relatively rare occurrence which should, however, be
kept in mind when considering the likelihood of association of
satellites that may pass all other criteria but are found to have
orbital planes approximately counter-parallel to the LMC.

\subsection{Contribution of LMC analogues to the primary satellite
  population}\label{SecLMCSatContrib}

\begin{figure}
\includegraphics[width=\linewidth]{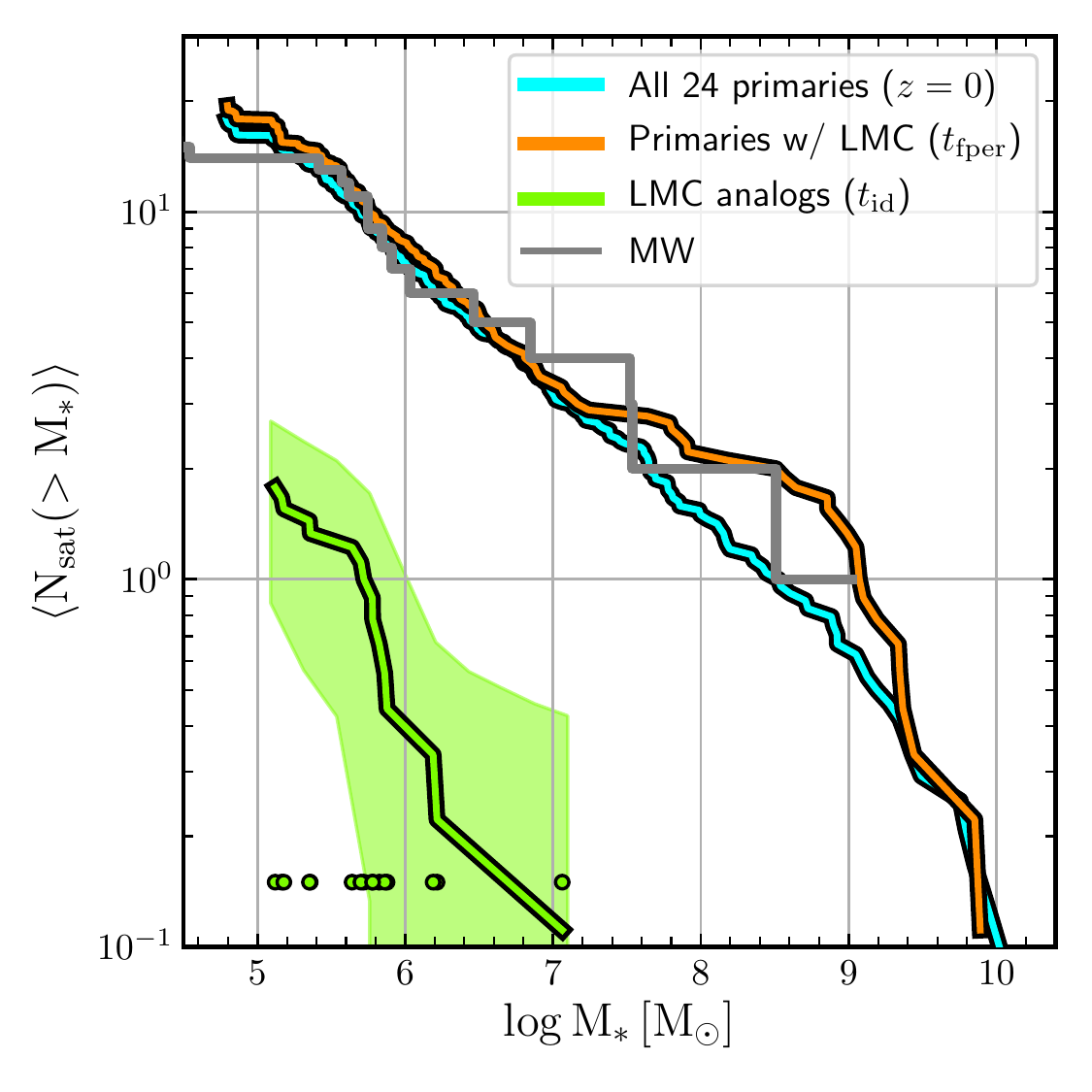}
\caption{Average satellite mass function for all the 24 primaries in
  AP-L2 runs at $z=0$ (cyan). This agrees fairly well with the observed
  satellite mass function in the MW (gray line). The satellite mass
  function of the $9$ primaries that contain a LMC analogue is shown in
  orange for comparison, and suggests an excess on the high-mass end
  due largely to the LMC analogue itself. On average, LMC analogues
  contribute 
  roughly 10\% of all satellites
  with $M_* > 10^5\; \rm M_\odot$ to their
  primaries (green curve). 
  The shaded area shows the
  $\pm1\sigma$ dispersion range.
  Green symbols show the individual masses of satellites
  identified in our $9$ LMC analogues. }
\label{fig:satmf}
\end{figure}

We consider now the contribution of satellites of LMC analogues to the
satellite population of the primary galaxy. The cyan curve in 
Fig.~\ref{fig:satmf} shows the average satellite mass function of all
$24$ APOSTLE primaries at $z=0$, and compares it to that of the $9$
primaries with LMC analogues (at the time of their first pericentric
passage; orange curve). 
Specifically, we consider all satellites within the virial radius
 of the primary ($\sim200$ kpc on average).
The grey curve shows the MW satellite 
population for reference (see Table~\ref{TabScores}). 
All MW satellites in our study are found within $\sim 250$ kpc of the MW centre, 
a distance that compares well with the virial radii of APOSTLE primaries.

The overall good agreement of APOSTLE with the MW satellite population
is reassuring, as it suggests that the simulated populations are
realistic and that their mass functions may be used to shed light on
the impact of the LMC on the overall MW satellite
population. Comparing the orange and cyan curves indicates that LMC
analogues have, as expected, a substantial impact on the massive end of
the satellite population, but, aside from that, the effect on the
whole population of satellites with $M_*>10^5\, M_\odot$ is relatively
modest. Indeed, the $9$ primaries with LMC analogues have $17.8^{+8.0}_{-1.2}$ 
(median and 25-75 percentiles)
such
satellites, compared with the average $16.1^{+6.3}_{-3.8}$ for all $24$ primaries
and with $16.3^{+4.6}_{-3.2}$ for the $15$ APOSTLE primaries without LMC
analogues. In other words, aside from the presence of the LMC itself,
the impact of the LMC satellites on the overall satellite population
is relatively minor.

This is also shown by the green curve in Fig.~\ref{fig:satmf}, which
indicates the (average) satellite mass function of the LMC analogues
at identification time, $t_{\rm id}$ (i.e., before infall). 
The $9$ LMC analogues contribute a total of $16$ dwarfs with
$M_*> 10^5\, M_\odot$ at infall, or roughly $\sim 10\%$ of the satellite
population of each primary. In terms of numbers, the average 
$\langle \rm N_{\rm sat}(M_{*}>10^5M_\odot)\rangle$
is $16/9=1.8\pm0.9$, where the error range specifies the $\pm 1\sigma$
spread of the distribution. 
The green circles at the bottom of Fig.~\ref{fig:satmf}
show the individual stellar masses of each satellite in our $9$ LMC
analogues. None of our LMC analogues has a companion as massive
as the SMC, which has a stellar mass of order
$M_*\sim3\times 10^8\, M_\odot$. Most satellites contributed by LMC
analogues have stellar masses $M_*<10^6\, M_\odot$.

We note that the relatively modest impact of the LMC on the MW massive
satellite population suggested by our results is consistent with the
early semi-analytical models of \citet{Dooley2017b}, as well as with
other studies of isolated LMC-mass systems using the FIRE simulations
\citep{Jahn2019} and simulations from the Auriga project
\citep{Pardy2020}.

\subsection{LMC and the radial distribution of satellites}
\label{ssec:raddist}
  
  \begin{figure}
\includegraphics[width=\linewidth]{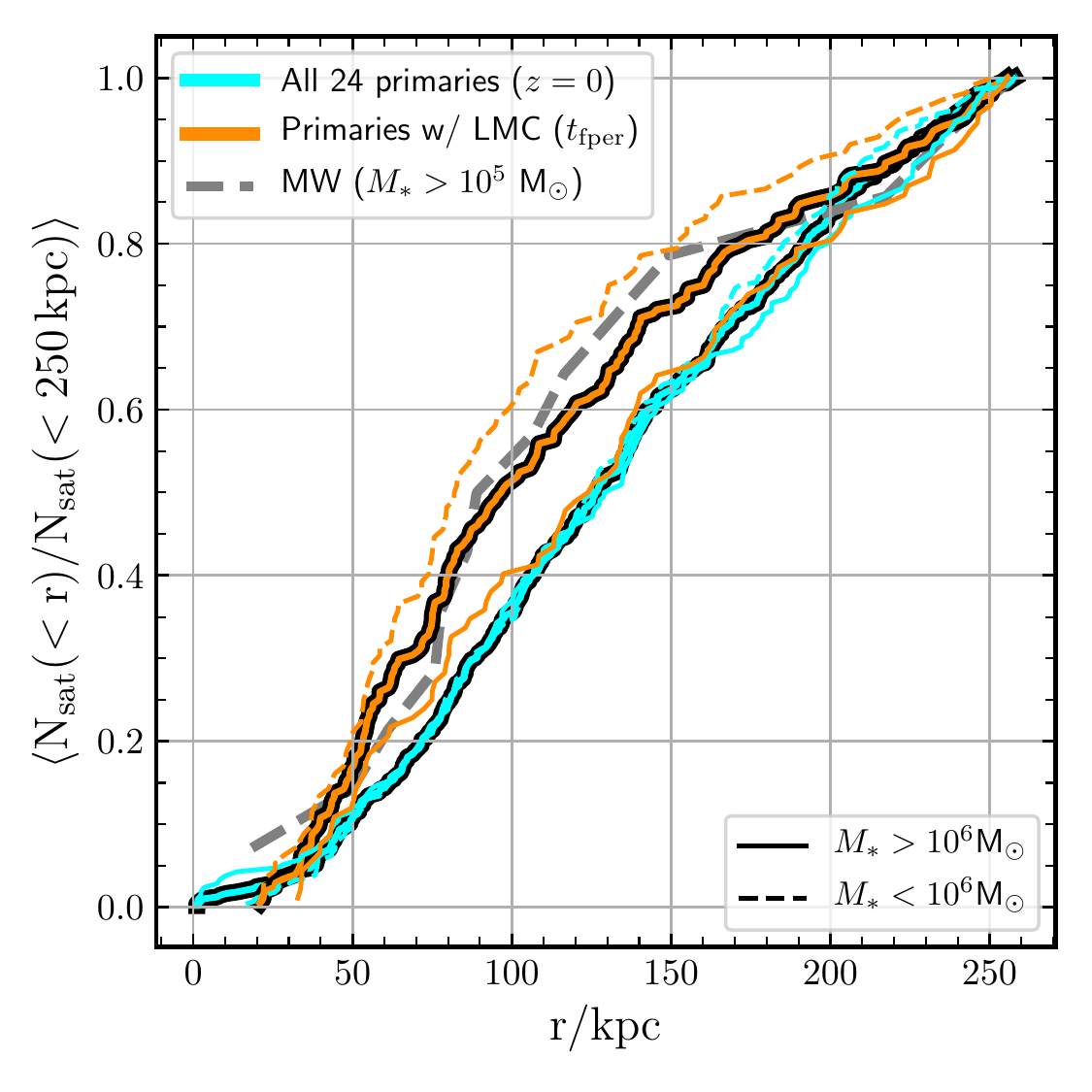}
\caption{Average cumulative radial distribution of satellites within
  250 kpc, for (i) all the 24 primaries in APOSTLE-L2, at $z=0$
  (cyan); (ii) the primaries of the 9 LMC analogues, at first pericentre
  (orange); and (iii) the Milky Way satellites. We include in all of
  these samples only satellites with $M_*>10^5\, M_\odot$. 
  Thinner lines show the distributions for simulated satellites filtered
  by stellar mass as quoted in the legend.  
  Note that
  the MW satellite distribution appears more concentrated than the
  average APOSTLE primary; this is well matched by systems with an LMC
  analogue, a transient configuration that results from the particular
  orbital configuration of the LMC
  and its satellites  
   (at pericentre).}
\label{fig:raddist}
\end{figure}
  
The radial distribution of satellites contains important clues to the
accretion history of a galaxy \citep[see; e.g.,][and references
therein]{Samuel2020,Carlsten2020}. Recent results from the SAGA survey
have suggested that 
\textit{"the radial distribution of MW satellites is much more concentrated
than the average distribution of SAGA satellites, mostly due to the presence of the 
LMC and SMC" \citep{Mao2020}.}
We explore below whether our simulations confirm that this
  effect is likely due to the LMC and its satellites.

The cyan curve in Fig.~\ref{fig:raddist} shows the average cumulative
radial distribution of all $M_*>10^5\, M_\odot$ satellites within
$250$ kpc of the $24$ APOSTLE primaries. The corresponding MW satellite population
is significantly more concentrated, as shown by the grey dashed curve
in the same figure\footnote{Radial distances for MW satellites
have been calculated from the RA, dec, $(m-M)$ data available in 
\citet{McConnachie2012}'s Nearby Dwarf Database (see references therein).}. 
Interestingly, the $9$ APOSTLE primaries with LMC
analogues, shown by the orange curve, also have more concentrated
satellite distributions, in good agreement with the MW satellite
population.

This is mainly a transient result of the particular
orbital phase of the LMC analogues, which are chosen to be near first
pericentric passage. Indeed, at $z=0$ the same $9$ primaries have less
centrally concentrated distributions, consistent with the average
result for all $24$ primaries (cyan curve). 
Support for our
interpretation of the transient concentration as due to the LMC
analogues and their associated satellite systems is provided by the
thin orange lines in Fig.~\ref{fig:raddist}. The dashed and solid
(thin) orange lines indicate results for systems with stellar mass
exceeding or smaller than $10^6\, M_\odot$. The higher concentration
is only apparent in the latter case: this is consistent with our
earlier finding that LMC analogues contribute mainly systems with
$M_*<10^6\, M_\odot$ (see Fig.~\ref{fig:satmf}).

We conclude that the concentrated radial distribution of satellites in
the Galaxy is probably a transient caused by the presence of the LMC
and its satellites near first pericentre. This transient effect
illustrates the importance of taking into account the particular
kinematic stage of the LMC when comparing the properties of the
Galactic satellite population with that of other external galaxies.

\begin{figure*}
\includegraphics[width=\linewidth]{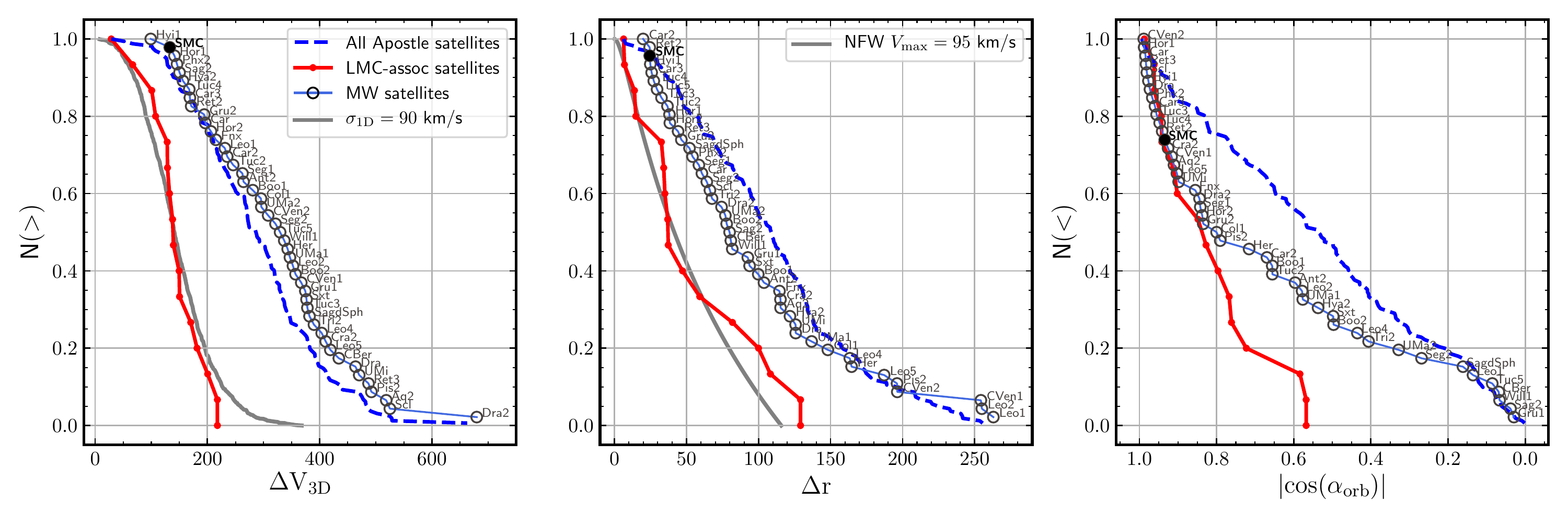}
\caption{Cumulative distributions of the three diagnostics used to
  rank MW satellites in terms of likely association with the
  LMC. These diagnostics are the 3D velocity relative to the LMC
  ($\Delta V_{\rm 3D}$, left), the radial distance relative to the LMC
  ($\Delta r$, centre) and the alignment with the LMC's orbital pole
  direction ($\vert \rm cos(\alpha_{\rm orb})\vert$, right). A red
  line shows the cumulative distribution for LMC-associated satellites
  in APOSTLE; a dashed blue line shows that for \textit{all} APOSTLE
  satellites, and a black line shows the distribution for all MW
  satellites, as labelled.  For reference, the grey curve in the
  $\Delta V_{\rm 3D}$ panel (left) shows a Gaussian distribution with
  $\sigma_{\rm 1D}=90$ km/s.  In the $\Delta r$ panel (centre) the
  grey curve shows the cumulative mass profile of an NFW dark matter
  halo with $V_{\rm max}=95$ km/s, roughly the average $V_{\rm max}$
  of the APOSTLE LMC analogues.  }
    \label{fig:diagnostics}
\end{figure*}

\section{LMC-associated satellites in the Milky Way}
\label{SecLMCIdCrit}

We have seen in the above subsections that satellites associated with
LMC-analogues contribute modestly to the primary satellite population,
and distinguish themselves from the rest of a primary's satellites by
their proximity in phase space to their parent LMC analogue. Satellites
closely aligned in orbital pole direction, and at small relative distances
and velocities from the LMC, should be strongly favoured in
any attempt to identify which MW satellites have been contributed by
the LMC.

We may compile a ranked list of potential associations by assigning to
all MW satellites numerical scores on each of the above
diagnostics. This score consists of a numerical value equal to the
fraction of associated satellites in the simulations that are farther
from their own LMC analogue in each particular dignostic (i.e., a score
of $1$ means that a particular satellite is closer to the LMC than
{\it all} simulated satellites, in that diagnostic.). We illustrate
this scoring procedure in Fig.~\ref{fig:diagnostics}.

The left panel shows the cumulative distribution of
$\Delta V_{\rm 3D}$, the relative velocity between the LMC and other
satellites. The red curve corresponds to all simulated satellites
associated to LMC analogues, the dashed blue curve to all satellites of
APOSTLE primaries. The grey curve shows the cumulative distribution
expected if associated satellites had a Gaussian isotropic velocity
distribution around the analogue with a velocity dispersion of $\sigma_{\rm 1D}=90$
km/s. For example, the SMC (highlighted in Fig.~\ref{fig:diagnostics}
with a filled circle) has $\Delta V_{\rm 3D}=133$ km/s, which gives it a
relatively high score of $\sim 0.59$ in this diagnostic. According to
this diagnostic, any MW satellite whose LMC relative velocity exceeds
$\sim 220$ km/s has a score of zero, and its association with the LMC
is in doubt.

\begin{figure*}
\includegraphics[width=\columnwidth]{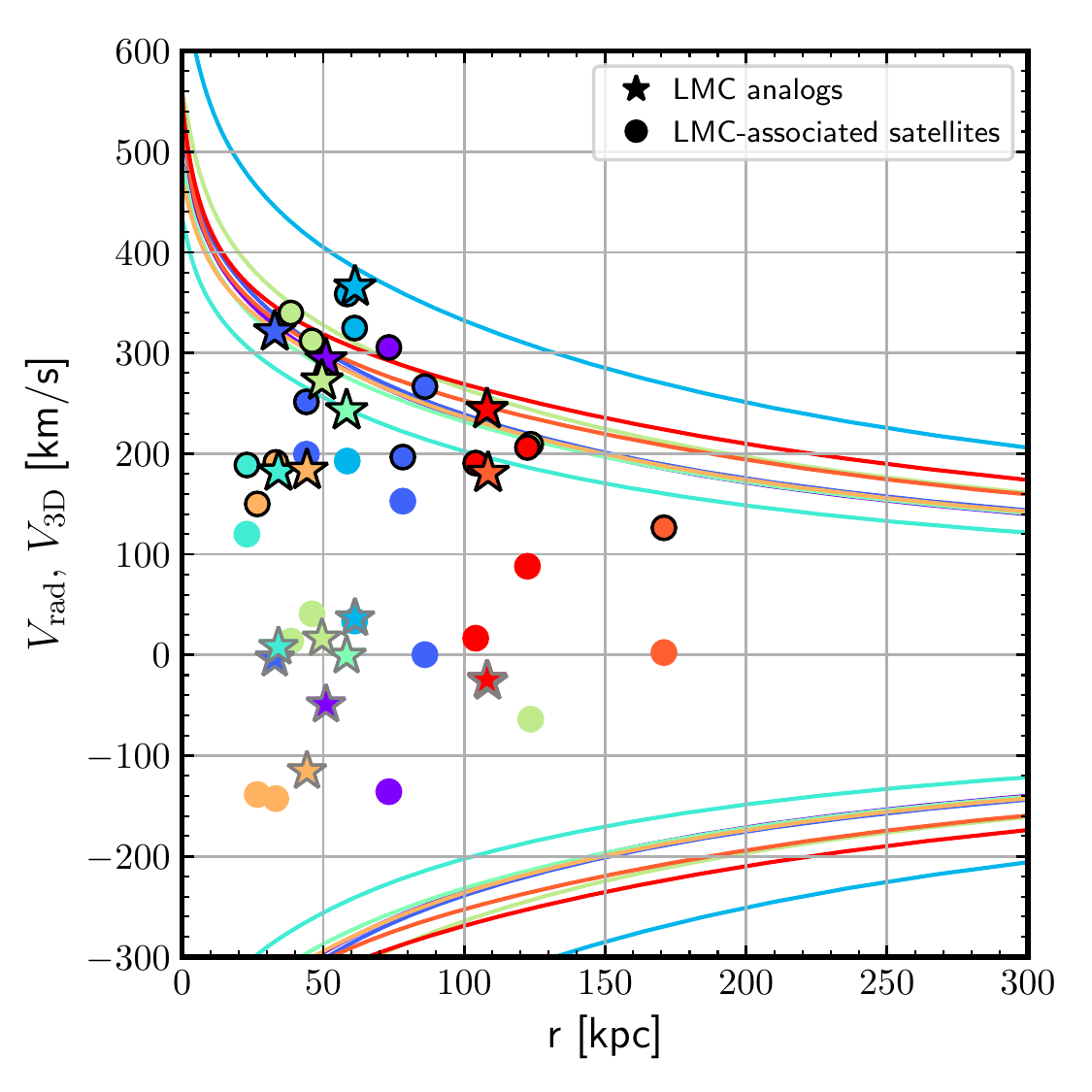}
\includegraphics[width=\columnwidth]{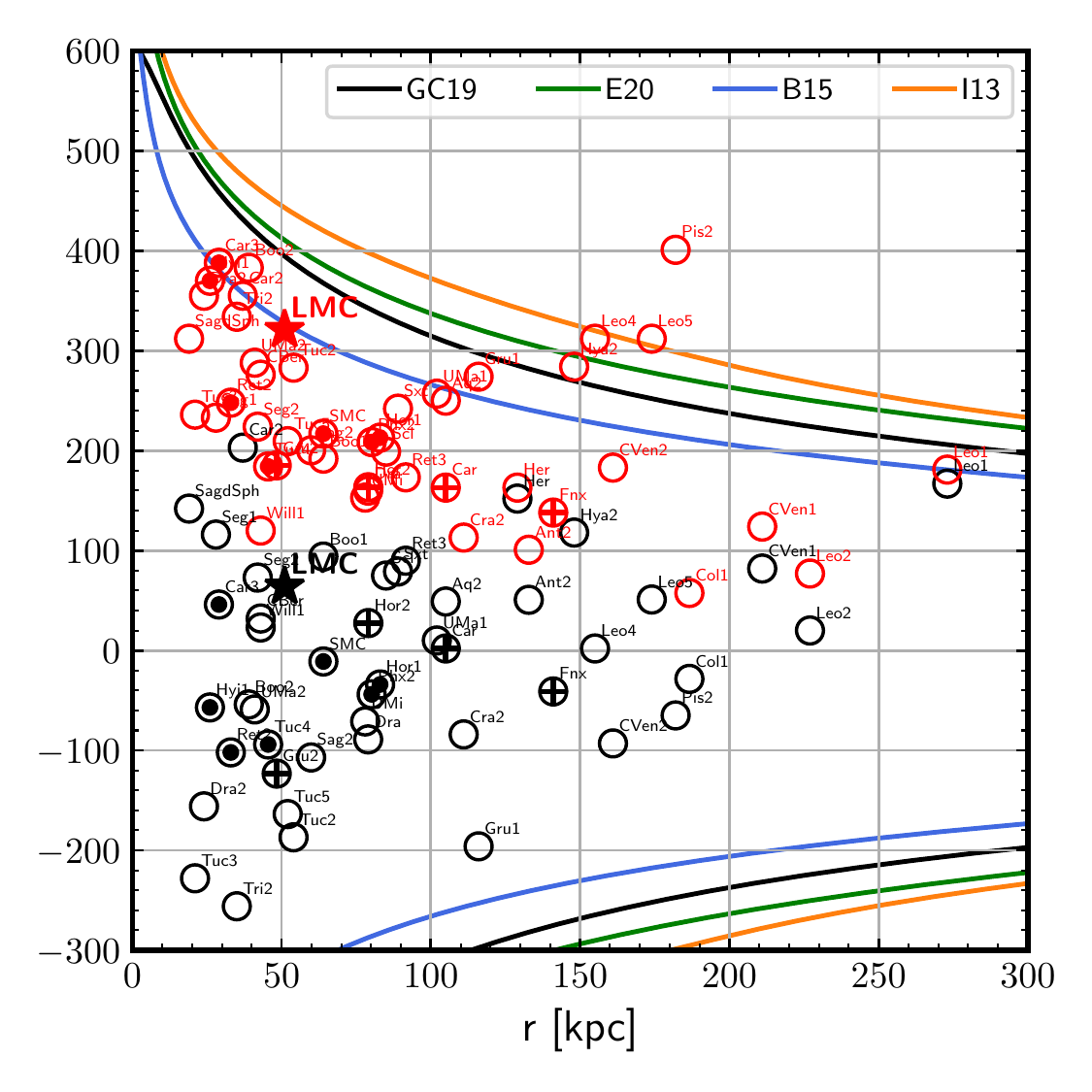}
\caption{Radial velocity $V_{\rm rad}$ and 3D velocity $V_{\rm 3D}$  versus  radial distance,  at pericentre.
  \textit{Left}: APOSTLE LMC analogues (stars) and LMC-associated
  satellites (circles). Radial (3D) velocities are shown with symbols
  with grey (black) edges. Lines illustrate the escape velocity
  profiles of the corresponding primaries. Color-coding is the same as
  in previous figures.  \textit{Right}: observed MW satellites at
  $z=0$. Radial velocities are shown in black, and 3D velocities in
  red. The LMC is marked with a star. Observed $V_{\rm rad}$ and
  $V_{\rm 3D}$ are from \citet{Fritz2018} when available, or computed
  from measured kinematic data as explained in Tab.~\ref{TabScores}.
  Lines show the escape velocity profiles derived from the following
  MW models proposed in the literature:
  \citet{GaravitoCamargo2019,Errani2020,Bovy2015,Irrgang2013}.  }
    \label{fig:vrv3d}
\end{figure*}

The middle and right panels of Fig.~\ref{fig:diagnostics} show the
other two diagnostics we have chosen to rank possible LMC-associated
satellites. The middle panel indicates the relative distance between
satellites and the LMC. The red curve again corresponds to simulated
satellites associated with LMC analogues. Its distribution is very well
approximated by the radial mass profile of an NFW halo with
$V_{\rm max}=90$ km/s and concentration $c=10.2$ (grey curve).  For
reference, the median $V_{\rm max}$ and 10-90 percentiles for LMC
analogues is $78^{+52}_{-16}$ km/s (see Fig.~\ref{fig:vmaxmstar}).
Together with the evidence from the left panel, this confirms that
satellites associated with LMC analogues are, at first pericentre,
distributed around the analogues more or less as they were before
infall. Tides, again, have not yet had time to disrupt the close
physical association of the Magellanic group in phase space. The SMC,
for example, scores $\sim 0.76$ in this diagnostic.

Finally, the right-hand panel of Fig.~\ref{fig:diagnostics} shows the
orbital pole alignment, where we have chosen to use the absolute value
of $\cos (\alpha_{\rm orb})$ in order to account for the possibility
of ``counter-rotating'' satellites. The SMC, again, scores high in
this diagnostic; with a score of $\sim 0.72$ for
$|\cos(\alpha_{\rm orb})|=0.93$. In this case, any MW satellites with
$|\cos(\alpha_{\rm orb})|<0.57$ would have a score of zero.

We may add up the three scores to rank all MW satellites according to
the likelihood of their association of the LMC. The data and scores
are listed in Table~\ref{TabScores}, and show that, out of $46$ MW
satellites, $11$ have non-zero scores in all three categories. Of
these $11$, the $7$ whose association appears firm are: Hydrus 1, SMC,
Car 3, Hor 1, Tuc 4, Ret 2, and Phx 2. These $7$ satellites are
highlighted with a solid central circle in the figures throughout the
paper. A second group with more tenuous association, mainly because of
their large relative velocity difference, contains Carina, Hor 2, and
Grus 2. The final member is Fornax, whose scores in relative velocity
and position are non-zero but quite marginal. These $4$ satellites are
highlighted with a cross in the figures.

Three satellites in this list have $M_*>10^5\, M_\odot$ (SMC, Carina,
Fornax).  This is actually in excellent agreement with the discussion
in Sec.~\ref{SecLMCSatContrib}, where we showed that LMC analogues bring
$\sim2$
 such satellites into their primaries.  The same arguments
suggest that $\sim 10\%$ of all MW satellites might have been
associated with the LMC. This small fraction is in tension with the
$11$ out of $46$ satellites (i.e., $24\%$) in our list. We note,
however, that our current list of MW satellites is likely very
incomplete \citep[see; e.g.,][]{Newton2018,Nadler2020}, and highly
biased to include more than its fair share of LMC satellites. Indeed,
many of the new satellite detections have been made possible by DES, a
survey of the southern sky in the vicinity of the Magellanic Clouds
\citep{Bechtol2015,Koposov15a,Drlica-Wagner2015}.

Our list adds some candidates compared to the lists compiled by
earlier work, but also contain some differences. \citet{Sales2011} and
\citet{Sales2017} identified only three satellites as clearly
associated with the LMC: the SMC, Hor 1 and Tuc 2. The latter is, however,
deemed unlikely given our analysis, especially because of its large
LMC relative velocity, $\Delta V_{\rm 3D}=246$ km/s.
\citet{Kallivayalil2018}'s list of possible LMC-associated satellites includes
Car 2, Draco 2, and Hydra 2. According to our analysis, the first two
are ruled out by their large relative velocity. The last one is, on the
other hand, ruled out by its large orbital pole deviation.

\citet{Erkal2020} claim SMC, Hydrus 1, Car 3, Hor 1, Car 2, Phx 2 and
Ret 2 as associated with the LMC.  Using a similar methodology,
\citet{Patel2020} also identifies the first 5 as LMC ``long term
companions''.  Of these, our analysis disfavours Car 2, again on
account of its large relative velocity, $\Delta V_{\rm 3D}=235$ km/s.  Finally,
\citet{Pardy2020} argues for Carina and Fornax as candidates for LMC
association. Our analysis does not rule out either (both have non-zero
scores in all three categories), although the evidence for association
is not particularly strong, especially for Fornax.  Our results agree
with \citet{Erkal2020} in this regard, who argues the need for an
uncommonly massive LMC to accomodate Fornax as one of its satellites.

\begin{figure*}
\includegraphics[width=\linewidth]{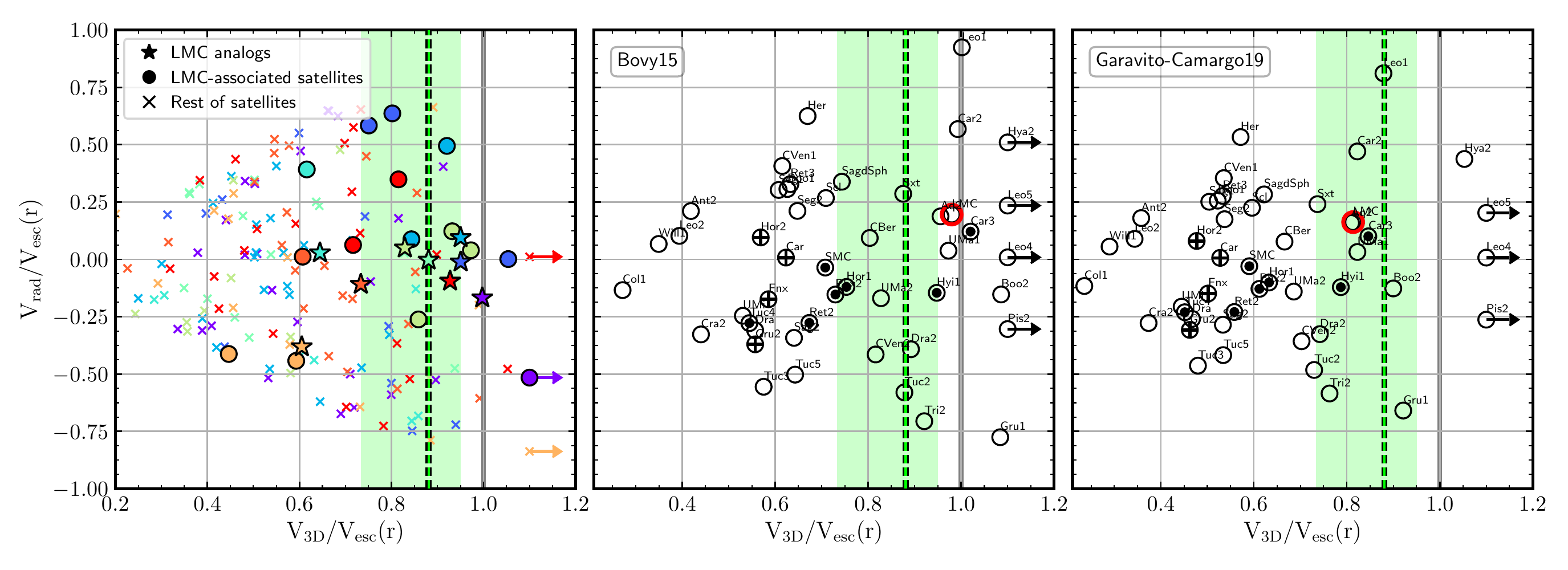}
\caption{ Radial velocity versus total 3D velocity, both normalized by
  the escape velocity at the pericentric radius.  \textit{Left}: APOSTLE LMC
  analogues, LMC-associated satellites, and rest of satellites of the
  corresponding primary at first pericentre. Color-coding is the same
  as in previous figures.  Centre: Observed MW satellites assuming the
  \citet{Bovy2015} MW potential.  \textit{Right}: Observed MW satellites
  assuming the \citet{GaravitoCamargo2019} MW potential.  A green
  vertical line with shade shows the median
  $V_{\rm 3D}/V_{\rm esc}(r)$ and 25-75\% percentiles for LMC analogues,
  and is marked in all panels.  In the centre and right panels the
  observed LMC's position, as defined by the assumed MW escape
  velocity, is highlighted with a red open circle.  Objects with
  $V_{\rm 3D}/V_{\rm esc}(r)>1$ are gravitationally unbound to the
  primary given that choice of potential. }
    \label{fig:vrv3d3pan}
\end{figure*}

\section{The LMC and the escape velocity of the Milky Way}
\label{SecVesc}

We have argued in the preceding sections that, because the LMC is
just past its first pericentric passage, then its associated
satellites must still be close in position and velocity. Other
corollaries are that both the LMC and its satellites must have
Galactocentric radial velocities much smaller than their tangential
velocities, and that their total velocities must approach the escape
velocity of the Milky Way at their location.

We explore this in the left panel of
Fig.~\ref{fig:vrv3d}, which shows the radial ($V_{\rm rad}$) and total
3D velocities ($V_{\rm 3D}$) of LMC analogues (stars) and LMC-associated
satellites (circles) at the LMC analogue's first pericentre, as a function of their
radial distance to the primary. Radial velocities are shown as symbols
without edges, and 3D velocities as symbols with dark edges. A
different color is used for each of the 9 LMC-analogue systems.

All LMC-analogues and most of their associated satellites are close to pericentre and have
therefore radial velocities much smaller than their total velocities:
half of the LMC analogues have $|V_{\rm rad}|/V_{\rm 3D}<0.10$, and half
of the $15$ associated satellites have
$|V_{\rm rad}|/V_{\rm 3D}<0.43$. (For reference, the LMC itself has
$|V_{\rm rad}|/V_{\rm 3D}\approx 0.2$.)

It is also clear from the left panel of Fig.~\ref{fig:vrv3d} that the
large majority of LMC analogues have total velocities that trace closely
the escape\footnote{Escape velocities are defined as the speed needed
  for a test particle to reach infinity, assuming spherical symmetry
  and that the mass of the primary halo does not extend beyond a
  radius $r=2\times r_{200}$.}  velocity of each of their primaries at
their location.  This is interesting because many commonly used models
for the MW potential are calibrated to match observations in and
around the solar circle, but differ in the outer regions of the
Galaxy, near the location of the LMC.

This is illustrated in the right-hand panel of Fig.~\ref{fig:vrv3d},
where the 4 different curves show the escape velocity curves
corresponding to models recently proposed for the Milky Way; i.e., those of
\citet[][I13]{Irrgang2013}, \citet[][B15]{Bovy2015}, \citet[][GC19]{GaravitoCamargo2019}, and
\citet[][E20]{Errani2020}. These models differ in their predicted escape
velocities at the location of the LMC ($r\sim 50$ kpc) from a low
value of $\sim 330$ km/s (B15) to a high value of $\sim 445$ km/s (I13). The LMC could
therefore provide useful additional information about the total virial mass
of the MW, which dominates any estimate of the escape velocity.

We explore this in more detail in the left panel of
Fig.~\ref{fig:vrv3d3pan}, where we show the radial and total velocity
of LMC analogues and their satellites, expressed in units of the
escape velocity at their current location.  The median
$V_{\rm 3D}/V_{\rm esc}$ and 25-75\% percentiles for LMC analogues is
$0.88^{+0.07}_{-0.15}$, a value that we indicate with a shaded green
line. For LMC-associated satellites the corresponding value is
similar; $0.82^{+0.11}_{-0.15}$, again highlighting the close
dynamical correspondence between LMC analogues and their satellites. The
high velocity of LMC-associated systems differs systematically from that
of regular satellites (i.e., those not associated with LMC analogues,
shown with colored crosses in Fig.~\ref{fig:vrv3d3pan}). These
systems have $V_{\rm 3D}/V_{\rm esc}=0.59^{+0.18}_{-0.14}$.

The well-defined value of $V_{\rm 3D}/V_{\rm esc}$ for LMC analogues allows us to
estimate the MW escape velocity at $50$ kpc from the total
Galactocentric velocity of the LMC, estimated at
$V_{\rm 3D} \approx 320$ km/s by \citet{Kallivayalil2013}. This
implies $V_{\rm esc}^{\rm MW}$(50 kpc)$\approx 365$ km/s, favouring
models with modest virial masses for the MW. 
The four models shown in the right-hand panel of Fig.~\ref{fig:vrv3d}
have $V_{\rm esc}(50\, \rm kpc)=$ 397 (GC19); 413 (E20);  330 (B15); 445 (I13) km/s.
Of these, the closest to our estimate is that of GC19, which has a
virial mass of $M_{200}=1.2\times10^{12}\, M_\odot$. Interestingly, this is also the
mass favored by the recent analysis of stellar halo kinematics by
\citet{Deason2020}.

Further constraints may be inferred by considering simulated
satellites with velocities higher than the local escape speed. These
are actually quite rare in our APOSTLE simulations: only $2$
LMC-associated satellites and $3$ regular satellites (out of a total
of $163$) appear ``unbound''. We compare this with observed MW
satellites in Fig.~\ref{fig:vrv3d3pan}, where the middle panel
corresponds to the B15 model potential and the right-hand panel to
that of GC19.  (MW satellite Galactocentric radial and 3D velocities
are taken from \citet{Fritz2018} if available, or otherwise computed
from measured kinematic data as explained in the caption to
Tab.~\ref{TabScores}.)  Although the LMC $V_{\rm 3D}/V_{\rm esc}$
seems acceptable in both cases, assuming the B15 potential would yield
$8$ escaping satellites out of $46$, a much higher fraction than
expected from the simulations. Even after removing Hya 2, Leo 4, Leo 5
and Pis 2, which are distant satellites with large velocity
uncertainties (in all these cases exceeding $\sim 250$ km/s), the
fraction of escapers would still be $\sim 10\%$, much larger than
predicted by APOSTLE.

The GC19 potential fares better, with three fewer escapers than B15:
Gru 1, Car 3 and Boo 2 are all comfortably bound in this
potential. Hya 2, Leo 4, Leo 5 and Pis 2 are still unbound,
however. Indeed, Leo 5 and Pis 2 would be unbound even in the I13
potential, the most massive of the four, with a virial mass
$M_{200}=1.9\times10^{12}\, M_\odot$. Should the velocities/distances of those
satellites hold, it is very difficult to see how to reconcile their
kinematics with our simulations, unless those velocities are
substantially overestimated. Tighter, more accurate estimates of their
kinematics should yield powerful constraints on the Galactic
potential.

\section{Summary and Conclusions}
\label{SecConc}

We have used the APOSTLE suite of cosmological hydrodynamical
simulations to study the accretion of 
LMC-mass satellites into the halo
of MW-sized galaxies.  APOSTLE consists of simulations of $12$
cosmological volumes selected to resemble the Local Group. Each volume
includes a pair of halos with halo masses, separation, and relative
radial and tangential velocities comparable to the MW and M31.  We
identify ``LMC analogues'' as massive satellites of any of the 24
APOSTLE primary galaxies. These satellites are chosen to be representative
of the recent accretion of the LMC into the Galactic halo, taking into
account the LMC stellar mass and its particular kinematic state near
the first pericentric passage of its orbit.

Our results allow us to address the role of the LMC (the most massive
Galactic satellite) on the properties of the MW satellite population,
including (i) the frequency of LMC-mass satellites around MW-sized
galaxies and the effects of the Local Group environment; (ii)
observational diagnostics of possible association between MW
satellites and the LMC before infall, (iii) the contribution of the
LMC to the population of ``classical'' satellites of the MW; and (iv)
the constraints on the MW gravitational potential provided by the LMC
motion. To our knowledge, this is the first study of ``LMC analogues''
and their satellite companions carried out in realistic Local Group 
cosmological hydrodynamical simulations.
 
Our main results may be summarized as follows. 

\begin{itemize}
 
\item We find that $14$ out of $24$ primaries in APOSTLE have a
  satellite of comparable mass to the LMC
  ($8.75 \leq \log M_*/M_\odot \leq 10$) within 350 kpc at $z=0$. This
  is a higher fraction than estimated in previous work. 
  We use the DOVE simulation to study the 
  frequency of massive satellites around MW-mass haloes that are isolated
  and in pairs.
  The high
  frequency of LMC analogues in APOSTLE seems to have an environmental
  origin, as 
  LMC-like companions are roughly twice more
  frequent around primaries in Local Group-like environments than
  around isolated halos of similar mass.

\item Out of the $14$ LMC analogues, we select a subsample of $9$ which
  have reached their first pericentric passage in the past $4$ Gyr.
  These satellites inhabit $M_{200}\sim10^{11}$ M$_\odot$ halos before
  infall, and have rather eccentric orbits, with median pericentric
  and apocentric 
  distances of  $\sim 60$ kpc and $\sim 420$ kpc, respectively.

\item LMC analogues host their own satellites and contribute them to the
  primary satellite population upon infall. We find a total of $16$
  LMC-associated satellites
  before infall  
   with $M_* > 10^5\; \rm M_\odot$ for the
  $9$ LMC analogues, or slightly fewer than $2$ ``classical'' satellites
  per LMC.
  One satellite merges with the LMC analogue before first pericentre.  
   The LMC satellites contribute, on average, about $\sim 10\%$
  of the total population of primary satellites.

\item In agreement with previous work, we find that at the time of
  first pericentre, LMC-associated satellites are all distributed
  close to, and along, the orbital plane of the LMC,
  extending over  $\sim 45^\circ$ along the leading and trailing part of
  the orbit. Their orbital angular momentum vectors are aligned with
  that of the LMC, with a median relative angle of
  $32^\circ$.
  
\item We report one case of an LMC-associated satellite that is
  apparently {\it counter-rotating} the primary compared with the
  LMC. The apparent counter-rotation may result when the orbital
  motion of the satellite around the LMC is comparable or larger than
  the pericentric distance of the LMC. Under some circumstances, this
  leads the satellite to approach the centre of the primary ``on the
  other side'' relative to the LMC. This is relatively rare, and only
  one of the $15$ LMC-associated satellites appears to
  ``counter-rotate''.

\item We find that LMC-associated satellites 
  are located very
  close to their LMC analogue in position and velocity, with a median
  relative radial distance of $\sim 37$ kpc and a median relative 3D
  velocitys of $\sim 138$ km/s. This is because there  has not
  been enough time for tidal interactions from the MW to disperse the
  original orbits of LMC-companion satellites.

\item We may use the proximity of  associated satellites to the LMC in
  phase space to rank MW satellites according to the likelihood of
  their LMC association. We find that $11$ out of $46$ MW satellites
  could in principle be LMC associates. For $7$ of those the
  association appears firm: Hydrus 1, SMC, Car 3, Hor 1, Tuc 4, Ret 2,
  and Phx 2. Others, such as Carina, Hor 2, Grus 2 and Fornax are
  potential associates as well, but their large LMC relative
  velocities weakens their case.
 
\item The radial distribution of the satellite populations of
  primaries with LMC analogues is more concentrated than those of
  average APOSTLE primaries. This effect is largely driven by the particular
  kinematic stage of the LMC, near its first pericentric passage, and
  largely disappears after the LMC (and its associated satellites)
  move away from pericentre. This offers a natural explanation  for the
  more concentrated radial distribution of satellites in the MW
  compared to observed MW-analogues in the field, as recently reported by the
  SAGA survey \citep{Mao2020}. 

\item The 3D velocity of LMC analogues near first pericentre is very
  close to the escape velocity of their primaries, with a median
  $V_{\rm 3D}/V_{\rm esc}\approx 0.9$. We may use this result to
  derive an estimate for the MW's escape velocity at the location of
  the LMC ($r\sim 50$ kpc) of $\sim 365$ km/s. We also find that very
  few simulated satellites (fewer than roughly $1$ in $30$) are unbound
  from their primaries.  This information may be used to discriminate
  between different models of the MW potential. We find the model
  proposed by \citet{GaravitoCamargo2019} to be
  in reasonable agreement with our constraints, suggesting a MW virial
  mass of roughly $1 \times 10^{12}\, M_\odot$.

\end{itemize}

Our analysis shows that $\Lambda$CDM simulations of the Local Group
can easily account for the properties of the Magellanic accretion into
the halo of the Milky Way, and offer simple diagnostics to guide the
interpretation of extant kinematic data when attempting to disentangle
Magellanic satellites from the satellite population of the Milky
Way. The accretion of the LMC and its associated satellites into the
Milky Way seems fully consistent with the hierarchical buildup of the
Galaxy expected in the $\Lambda$CDM paradigm of structure formation.

\begin{table*}
\centering
\small
\caption{Values and 'scores' of MW satellites according to the
  different diagnostics used in this paper to assess association with
  the LMC: the 3D velocity relative to the LMC ($\Delta V_{\rm 3D}$), the
  radial distance relative to the LMC ($\Delta r$), and the alignment
  with the LMC's orbital pole direction
  ($|\rm cos(\alpha_{\rm orb})|$).  MW satellites are ordered
  according to their \textit{total} score in these 3 categories (last
  column).  The 11 MW satellites which we consider in this paper may
  be possibly associated to the LMC according to APOSTLE predictions
  (i.e., those with non-zero scores in all 3 categories) are
  highlighted in red.  Column 3 indicates if the satellite is
  co-rotating (+) or counter-rotating (-) the primary with respect to
  the LMC.  Column 4 shows the stellar mass of MW satellites, computed
  applying a mass-to-light ratio to the $V$-band luminosities in
  \citet{McConnachie2012}'s database. We assume $M_*/L_V=1.6$ for all
  satellites (appropriate for dSph-type galaxies) except for the SMC,
  where $M_*/L_V=0.7$ has been used \citep[see][]{Woo2008}.  We
  consider all MW satellites for which kinematic data is available.
  For all satellites we adopt the positions and distance moduli
  data (RA, dec, $(m-M)$) in \citet{McConnachie2012}'s database.
  Line-of-sight velocities and proper motions have been taken from
  \citet{Fritz2018} (their Table~2) when available, and otherwise from
  \citet{McConnachie2020} (Tables~1 and ~4). SMC kinematic data is from
  \citet{Kallivayalil2013}.  Galactocentric positions and velocities
  have been computed assuming a distance of the Sun from the Milky Way
  of $R_\odot=8.29$ kpc, a circular velocity of the local standard of
  rest (LSR) of $V_0=239$ km/s \citep{McMillan2011}, and a peculiar
  velocity of the Sun with respect to the LSR of
  $(U_\odot,V_\odot,W_\odot) = (11.1, 12.24,7.25)$ km/s
  \citep{Schonrich2010}.  }

\begin{tabular}{ l l l l l l l l l l l }
\toprule
MW satellite & & Sign & $M_*$/($10^5$M$_\odot$) & $\Delta V_{\rm 3D}$/km s$^{-1}$& $\Delta r$/kpc & $\rm |cos(\alpha_{\rm orb})|$ & Score $\Delta V_{\rm 3D}$& Score $\Delta r$ & Score $|\rm cos(\alpha_{\rm orb})|$ & Total Score   \\

\midrule
\midrule
\red{Hydrus1 }   &    Hyi1    &    +    &    0.10    &    99.01    &    24.87    &    0.98    &    0.87    &    0.76    &    0.97    &    2.60 \\
\red{SMC}    &    SMC    &    +    &    3229.22    &    132.97    &    24.47    &    0.93    &    0.59    &    0.76    &    0.72    &    2.07 \\
\red{Horologium1}    &    Hor1    &    +    &    0.04    &    141.33    &    38.19    &    0.99    &    0.45    &    0.46    &    1.00    &    1.91 \\
\red{Carina3 }   &    Car3    &    +    &    0.01    &    168.75    &    25.81    &    0.96    &    0.27    &    0.76    &    0.86    &    1.89 \\
\red{Tucana4}    &    Tuc4    &    +    &    0.03    &    167.70    &    27.29    &    0.95    &    0.28    &    0.75    &    0.82    &    1.84 \\
\red{Reticulum2 }   &    Ret2    &    +    &    0.05    &    171.09    &    24.43    &    0.94    &    0.26    &    0.76    &    0.73    &    1.76 \\
\red{Phoenix2}    &    Phx2    &    +    &    0.03    &    145.83    &    54.18    &    0.97    &    0.42    &    0.36    &    0.95    &    1.73 \\
Tucana3    &    Tuc3    &    -    &    0.01    &    378.12    &    32.64    &    0.96    &    0.00    &    0.73    &    0.84    &    1.57 \\
\red{Carina}    &    Car    &    +    &    8.09    &    196.22    &    60.69    &    0.98    &    0.15    &    0.33    &    0.99    &    1.47 \\
Reticulum3    &    Ret3    &    -    &    0.03    &    487.03    &    44.15    &    0.98    &    0.00    &    0.42    &    0.99    &    1.41 \\
Sculptor    &    Scl    &    -    &    29.12    &    525.06    &    66.25    &    0.98    &    0.00    &    0.31    &    0.98    &    1.30 \\
\red{Horologium2}    &    Hor2    &    +    &    0.01    &    206.19    &    38.43    &    0.84    &    0.11    &    0.46    &    0.50    &    1.07 \\
\red{Grus2}    &    Gru2    &    +    &    0.05    &    194.33    &    46.43    &    0.83    &    0.15    &    0.41    &    0.49    &    1.05 \\
Draco    &    Dra    &    +    &    4.17    &    463.86    &    125.79    &    0.97    &    0.00    &    0.08    &    0.96    &    1.04 \\
CanesVenatici2    &    CVen2    &    -    &    0.16    &    308.17    &    196.32    &    0.99    &    0.00    &    0.00    &    1.00    &    1.00 \\
Carina2    &    Car2    &    +    &    0.09    &    235.30    &    19.87    &    0.67    &    0.00    &    0.78    &    0.17    &    0.96 \\
Segue1    &    Seg1    &    -    &    0.00    &    262.35    &    58.59    &    0.84    &    0.00    &    0.34    &    0.51    &    0.85 \\
Draco2    &    Dra2    &    +    &    0.02    &    679.81    &    74.32    &    0.84    &    0.00    &    0.29    &    0.52    &    0.81 \\
Crater2    &    Cra2    &    +    &    2.61    &    410.77    &    115.22    &    0.93    &    0.00    &    0.11    &    0.70    &    0.81 \\
Aquarius2    &    Aq2    &    -    &    0.08    &    518.58    &    115.28    &    0.91    &    0.00    &    0.11    &    0.66    &    0.77 \\
Tucana5    &    Tuc5    &    +    &    0.01    &    329.49    &    29.56    &    0.09    &    0.00    &    0.74    &    0.00    &    0.74 \\
\red{Fornax}    &    Fnx    &    +    &    331.22    &    215.01    &    114.53    &    0.86    &    0.08    &    0.11    &    0.54    &    0.73 \\
Tucana2    &    Tuc2    &    +    &    0.05    &    245.89    &    36.80    &    0.66    &    0.00    &    0.54    &    0.17    &    0.71 \\
CanesVenatici1    &    CVen1    &    +    &    3.73    &    367.17    &    254.35    &    0.92    &    0.00    &    0.00    &    0.68    &    0.68 \\
UrsaMinor    &    UMi    &    +    &    5.60    &    470.43    &    125.73    &    0.90    &    0.00    &    0.08    &    0.60    &    0.67 \\
Leo5    &    Leo5    &    -    &    0.08    &    419.04    &    187.19    &    0.90    &    0.00    &    0.00    &    0.61    &    0.61 \\
Sagittarius2    &    Sag2    &    +    &    0.17    &    150.34    &    79.92    &    0.04    &    0.33    &    0.27    &    0.00    &    0.61 \\
Columba1    &    Col1    &    +    &    0.09    &    295.47    &    148.11    &    0.80    &    0.00    &    0.00    &    0.41    &    0.41 \\
Hydra2    &    Hya2    &    +    &    0.09    &    156.49    &    121.81    &    0.54    &    0.31    &    0.09    &    0.00    &    0.40 \\
Pisces2    &    Pis2    &    +    &    0.07    &    492.15    &    196.14    &    0.79    &    0.00    &    0.00    &    0.39    &    0.39 \\
SagittariusdSph    &    SagdSph    &    -    &    343.65    &    381.88    &    52.08    &    0.16    &    0.00    &    0.37    &    0.00    &    0.37 \\
Bootes1    &    Boo1    &    -    &    0.35    &    280.81    &    99.81    &    0.66    &    0.00    &    0.20    &    0.17    &    0.37 \\
Segue2    &    Seg2    &    -    &    0.01    &    321.30    &    64.08    &    0.27    &    0.00    &    0.32    &    0.00    &    0.32 \\
Antlia2    &    Ant2    &    +    &    5.60    &    264.32    &    103.77    &    0.60    &    0.00    &    0.17    &    0.14    &    0.31 \\
Triangulum2    &    Tri2    &    -    &    0.01    &    389.81    &    67.73    &    0.41    &    0.00    &    0.31    &    0.00    &    0.31 \\
UrsaMajor2    &    UMa2    &    -    &    0.07    &    296.04    &    76.99    &    0.33    &    0.00    &    0.28    &    0.00    &    0.28 \\
Bootes2    &    Boo2    &    -    &    0.02    &    357.00    &    77.87    &    0.50    &    0.00    &    0.28    &    0.00    &    0.28 \\
ComaBerenices    &    CBer    &    +    &    0.08    &    434.12    &    80.77    &    0.07    &    0.00    &    0.27    &    0.00    &    0.27 \\
Willman1    &    Will1    &    +    &    0.01    &    336.92    &    81.77    &    0.07    &    0.00    &    0.27    &    0.00    &    0.27 \\
Grus1    &    Gru1    &    +    &    0.03    &    374.43    &    92.55    &    0.03    &    0.00    &    0.23    &    0.00    &    0.23 \\
Sextans    &    Sxt    &    +    &    6.98    &    376.57    &    93.79    &    0.50    &    0.00    &    0.22    &    0.00    &    0.22 \\
Hercules    &    Her    &    -    &    0.29    &    342.98    &    164.59    &    0.72    &    0.00    &    0.00    &    0.20    &    0.20 \\
Leo2    &    Leo2    &    +    &    10.77    &    352.40    &    255.00    &    0.58    &    0.00    &    0.00    &    0.11    &    0.11 \\
UrsaMajor1    &    UMa1    &    -    &    0.15    &    347.02    &    136.85    &    0.58    &    0.00    &    0.00    &    0.10    &    0.10 \\
Leo1    &    Leo1    &    +    &    70.49    &    231.14    &    262.99    &    0.14    &    0.00    &    0.00    &    0.00    &    0.00 \\
Leo4    &    Leo4    &    -    &    0.14    &    403.60    &    163.35    &    0.43    &    0.00    &    0.00    &    0.00    &    0.00 \\
\bottomrule
\end{tabular}
\label{TabScores}
\end{table*}

\section*{Data availability}
The simulation data underlying this article can be shared on reasonable
request to the corresponding author.
The observational data for Milky Way satellites used in this article comes from the following references:
 \citet[][see \url{http://www.astro.uvic.ca/~alan/Nearby_Dwarf_Database_files/NearbyGalaxies.dat}, and references therein]{Kallivayalil2013,Fritz2018,McConnachie2020,McConnachie2012}.

\section*{Acknowledgements}
We wish to acknowledge the generous contributions of all those who made 
possible the Virgo Consortium’s EAGLE/APOSTLE and DOVE simulation projects.
 ISS is supported by the Arthur B. McDonald Canadian
Astroparticle Physics Research Institute. JFN is a Fellow of the
Canadian Institute for Advanced Research.
AF acknowledges support by the Science and Technology Facilities Council (STFC)
 [grant number ST/P000541/1] and the Leverhulme Trust.
LVS is thankful for financial support from the Hellman Foundation as well
as NSF and NASA grants, AST-1817233 and HST-AR-14552. 
 This work used the DiRAC@Durham facility
managed by the Institute for Computational Cosmology on behalf of the
STFC DiRAC HPC Facility (www.dirac.ac.uk). The equipment was funded by
BEIS capital funding via STFC capital grants ST/K00042X/1,
ST/P002293/1, ST/R002371/1 and ST/S002502/1, Durham University and
STFC operations grant ST/R000832/1. DiRAC is part of the National
e-Infrastructure. 

\bibliographystyle{mnras}
\bibliography{archive}

\begin{thebibliography}{}
\makeatletter
\relax
\def\mn@urlcharsother{\let\do\@makeother \do\$\do\&\do\#\do\^\do\_\do\%\do\~}
\def\mn@doi{\begingroup\mn@urlcharsother \@ifnextchar [ {\mn@doi@}
  {\mn@doi@[]}}
\def\mn@doi@[#1]#2{\def\@tempa{#1}\ifx\@tempa\@empty \href
  {http://dx.doi.org/#2} {doi:#2}\else \href {http://dx.doi.org/#2} {#1}\fi
  \endgroup}
\def\mn@eprint#1#2{\mn@eprint@#1:#2::\@nil}
\def\mn@eprint@arXiv#1{\href {http://arxiv.org/abs/#1} {{\tt arXiv:#1}}}
\def\mn@eprint@dblp#1{\href {http://dblp.uni-trier.de/rec/bibtex/#1.xml}
  {dblp:#1}}
\def\mn@eprint@#1:#2:#3:#4\@nil{\def\@tempa {#1}\def\@tempb {#2}\def\@tempc
  {#3}\ifx \@tempc \@empty \let \@tempc \@tempb \let \@tempb \@tempa \fi \ifx
  \@tempb \@empty \def\@tempb {arXiv}\fi \@ifundefined
  {mn@eprint@\@tempb}{\@tempb:\@tempc}{\expandafter \expandafter \csname
  mn@eprint@\@tempb\endcsname \expandafter{\@tempc}}}

\bibitem[\protect\citeauthoryear{{Bechtol} et~al.,}{{Bechtol}
  et~al.}{2015}]{Bechtol2015}
{Bechtol} K.,  et~al., 2015, \mn@doi [\apj] {10.1088/0004-637X/807/1/50}, \href
  {https://ui.adsabs.harvard.edu/abs/2015ApJ...807...50B} {807, 50}

\bibitem[\protect\citeauthoryear{{Behroozi}, {Wechsler}  \&
  {Conroy}}{{Behroozi} et~al.}{2013}]{Behroozi2013}
{Behroozi} P.~S.,  {Wechsler} R.~H.,   {Conroy} C.,  2013, \mn@doi [\apj]
  {10.1088/0004-637X/770/1/57}, \href
  {http://adsabs.harvard.edu/abs/2013ApJ...770...57B} {770, 57}

\bibitem[\protect\citeauthoryear{{Besla}, {Kallivayalil}, {Hernquist},
  {Robertson}, {Cox}, {van der Marel}  \& {Alcock}}{{Besla}
  et~al.}{2007}]{Besla2007}
{Besla} G.,  {Kallivayalil} N.,  {Hernquist} L.,  {Robertson} B.,  {Cox} T.~J.,
   {van der Marel} R.~P.,   {Alcock} C.,  2007, \mn@doi [\apj]
  {10.1086/521385}, \href
  {https://ui.adsabs.harvard.edu/abs/2007ApJ...668..949B} {668, 949}

\bibitem[\protect\citeauthoryear{{Bovy}}{{Bovy}}{2015}]{Bovy2015}
{Bovy} J.,  2015, \mn@doi [\apjs] {10.1088/0067-0049/216/2/29}, \href
  {https://ui.adsabs.harvard.edu/abs/2015ApJS..216...29B} {216, 29}

\bibitem[\protect\citeauthoryear{{Boylan-Kolchin}, {Springel}, {White}  \&
  {Jenkins}}{{Boylan-Kolchin} et~al.}{2010}]{Boylan-Kolchin2010}
{Boylan-Kolchin} M.,  {Springel} V.,  {White} S. D.~M.,   {Jenkins} A.,  2010,
  \mn@doi [\mnras] {10.1111/j.1365-2966.2010.16774.x}, \href
  {https://ui.adsabs.harvard.edu/abs/2010MNRAS.406..896B} {406, 896}

\bibitem[\protect\citeauthoryear{{Boylan-Kolchin}, {Besla}  \&
  {Hernquist}}{{Boylan-Kolchin} et~al.}{2011}]{BoylanKolchin2011a}
{Boylan-Kolchin} M.,  {Besla} G.,   {Hernquist} L.,  2011, \mn@doi [\mnras]
  {10.1111/j.1365-2966.2011.18495.x}, \href
  {https://ui.adsabs.harvard.edu/abs/2011MNRAS.414.1560B} {414, 1560}

\bibitem[\protect\citeauthoryear{{Busha}, {Marshall}, {Wechsler}, {Klypin}  \&
  {Primack}}{{Busha} et~al.}{2011}]{Busha2011}
{Busha} M.~T.,  {Marshall} P.~J.,  {Wechsler} R.~H.,  {Klypin} A.,   {Primack}
  J.,  2011, \mn@doi [\apj] {10.1088/0004-637X/743/1/40}, \href
  {https://ui.adsabs.harvard.edu/abs/2011ApJ...743...40B} {743, 40}

\bibitem[\protect\citeauthoryear{{Campbell} et~al.,}{{Campbell}
  et~al.}{2017}]{Campbell2017}
{Campbell} D. J.~R.,  et~al., 2017, \mn@doi [\mnras] {10.1093/mnras/stx975},
  \href {https://ui.adsabs.harvard.edu/abs/2017MNRAS.469.2335C} {469, 2335}

\bibitem[\protect\citeauthoryear{{Carlsten}, {Greene}, {Peter}, {Greco}  \&
  {Beaton}}{{Carlsten} et~al.}{2020}]{Carlsten2020}
{Carlsten} S.~G.,  {Greene} J.~E.,  {Peter} A. H.~G.,  {Greco} J.~P.,
  {Beaton} R.~L.,  2020, \mn@doi [\apj] {10.3847/1538-4357/abb60b}, \href
  {https://ui.adsabs.harvard.edu/abs/2020ApJ...902..124C} {902, 124}

\bibitem[\protect\citeauthoryear{{Crain} et~al.,}{{Crain}
  et~al.}{2015}]{Crain2015}
{Crain} R.~A.,  et~al., 2015, \mn@doi [\mnras] {10.1093/mnras/stv725}, \href
  {https://ui.adsabs.harvard.edu/abs/2015MNRAS.450.1937C} {450, 1937}

\bibitem[\protect\citeauthoryear{{D'Onghia} \& {Fox}}{{D'Onghia} \&
  {Fox}}{2016}]{DOnghia2016}
{D'Onghia} E.,  {Fox} A.~J.,  2016, \mn@doi [\araa]
  {10.1146/annurev-astro-081915-023251}, \href
  {https://ui.adsabs.harvard.edu/abs/2016ARA&A..54..363D} {54, 363}

\bibitem[\protect\citeauthoryear{{D'Onghia} \& {Lake}}{{D'Onghia} \&
  {Lake}}{2008}]{DOnghia2008}
{D'Onghia} E.,  {Lake} G.,  2008, \mn@doi [\apjl] {10.1086/592995}, \href
  {https://ui.adsabs.harvard.edu/abs/2008ApJ...686L..61D} {686, L61}

\bibitem[\protect\citeauthoryear{{Davis}, {Efstathiou}, {Frenk}  \&
  {White}}{{Davis} et~al.}{1985}]{Davis1985}
{Davis} M.,  {Efstathiou} G.,  {Frenk} C.~S.,   {White} S.~D.~M.,  1985,
  \mn@doi [\apj] {10.1086/163168}, \href
  {http://adsabs.harvard.edu/cgi-bin/nph-bib_query?bibcode=1985ApJ...292..371D&db_key=AST}
  {292, 371}

\bibitem[\protect\citeauthoryear{{Deason} et~al.,}{{Deason}
  et~al.}{2020}]{Deason2020}
{Deason} A.~J.,  et~al., 2020, arXiv e-prints, \href
  {https://ui.adsabs.harvard.edu/abs/2020arXiv201013801D} {p. arXiv:2010.13801}

\bibitem[\protect\citeauthoryear{{Dooley}, {Peter}, {Carlin}, {Frebel},
  {Bechtol}  \& {Willman}}{{Dooley} et~al.}{2017}]{Dooley2017b}
{Dooley} G.~A.,  {Peter} A. H.~G.,  {Carlin} J.~L.,  {Frebel} A.,  {Bechtol}
  K.,   {Willman} B.,  2017, \mn@doi [\mnras] {10.1093/mnras/stx2001}, \href
  {https://ui.adsabs.harvard.edu/abs/2017MNRAS.472.1060D} {472, 1060}

\bibitem[\protect\citeauthoryear{{Drlica-Wagner} et~al.,}{{Drlica-Wagner}
  et~al.}{2015}]{Drlica-Wagner2015}
{Drlica-Wagner} A.,  et~al., 2015, \mn@doi [\apj]
  {10.1088/0004-637X/813/2/109}, \href
  {https://ui.adsabs.harvard.edu/abs/2015ApJ...813..109D} {813, 109}

\bibitem[\protect\citeauthoryear{{Erkal} \& {Belokurov}}{{Erkal} \&
  {Belokurov}}{2020}]{Erkal2020}
{Erkal} D.,  {Belokurov} V.~A.,  2020, \mn@doi [\mnras]
  {10.1093/mnras/staa1238}, \href
  {https://ui.adsabs.harvard.edu/abs/2020MNRAS.495.2554E} {495, 2554}

\bibitem[\protect\citeauthoryear{{Errani} \& {Pe{\~n}arrubia}}{{Errani} \&
  {Pe{\~n}arrubia}}{2020}]{Errani2020}
{Errani} R.,  {Pe{\~n}arrubia} J.,  2020, \mn@doi [\mnras]
  {10.1093/mnras/stz3349}, \href
  {https://ui.adsabs.harvard.edu/abs/2020MNRAS.491.4591E} {491, 4591}

\bibitem[\protect\citeauthoryear{{Fattahi} et~al.,}{{Fattahi}
  et~al.}{2016}]{Fattahi2016}
{Fattahi} A.,  et~al., 2016, \mn@doi [\mnras] {10.1093/mnras/stv2970}, \href
  {https://ui.adsabs.harvard.edu/abs/2016MNRAS.457..844F} {457, 844}

\bibitem[\protect\citeauthoryear{{Fattahi}, {Navarro}, {Frenk}, {Oman},
  {Sawala}  \& {Schaller}}{{Fattahi} et~al.}{2018}]{Fattahi2018}
{Fattahi} A.,  {Navarro} J.~F.,  {Frenk} C.~S.,  {Oman} K.~A.,  {Sawala} T.,
  {Schaller} M.,  2018, \mn@doi [\mnras] {10.1093/mnras/sty408}, \href
  {https://ui.adsabs.harvard.edu/abs/2018MNRAS.476.3816F} {476, 3816}

\bibitem[\protect\citeauthoryear{{Fritz}, {Battaglia}, {Pawlowski},
  {Kallivayalil}, {van der Marel}, {Sohn}, {Brook}  \& {Besla}}{{Fritz}
  et~al.}{2018}]{Fritz2018}
{Fritz} T.~K.,  {Battaglia} G.,  {Pawlowski} M.~S.,  {Kallivayalil} N.,  {van
  der Marel} R.,  {Sohn} S.~T.,  {Brook} C.,   {Besla} G.,  2018, \mn@doi
  [\aap] {10.1051/0004-6361/201833343}, \href
  {https://ui.adsabs.harvard.edu/abs/2018A&A...619A.103F} {619, A103}

\bibitem[\protect\citeauthoryear{{Gaia Collaboration} et~al.,}{{Gaia
  Collaboration} et~al.}{2018}]{GaiaColl2018}
{Gaia Collaboration} et~al., 2018, \mn@doi [\aap]
  {10.1051/0004-6361/201832698}, \href
  {https://ui.adsabs.harvard.edu/abs/2018A&A...616A..12G} {616, A12}

\bibitem[\protect\citeauthoryear{{Garavito-Camargo}, {Besla}, {Laporte},
  {Johnston}, {G{\'o}mez}  \& {Watkins}}{{Garavito-Camargo}
  et~al.}{2019}]{GaravitoCamargo2019}
{Garavito-Camargo} N.,  {Besla} G.,  {Laporte} C. F.~P.,  {Johnston} K.~V.,
  {G{\'o}mez} F.~A.,   {Watkins} L.~L.,  2019, \mn@doi [\apj]
  {10.3847/1538-4357/ab32eb}, \href
  {https://ui.adsabs.harvard.edu/abs/2019ApJ...884...51G} {884, 51}

\bibitem[\protect\citeauthoryear{{Garrison-Kimmel}, {Boylan-Kolchin}, {Bullock}
   \& {Lee}}{{Garrison-Kimmel} et~al.}{2014}]{Garrison-Kimmel2014}
{Garrison-Kimmel} S.,  {Boylan-Kolchin} M.,  {Bullock} J.~S.,   {Lee} K.,
  2014, \mn@doi [\mnras] {10.1093/mnras/stt2377}, \href
  {https://ui.adsabs.harvard.edu/abs/2014MNRAS.438.2578G} {438, 2578}

\bibitem[\protect\citeauthoryear{{Irrgang}, {Wilcox}, {Tucker}  \&
  {Schiefelbein}}{{Irrgang} et~al.}{2013}]{Irrgang2013}
{Irrgang} A.,  {Wilcox} B.,  {Tucker} E.,   {Schiefelbein} L.,  2013, \mn@doi
  [\aap] {10.1051/0004-6361/201220540}, \href
  {https://ui.adsabs.harvard.edu/abs/2013A&A...549A.137I} {549, A137}

\bibitem[\protect\citeauthoryear{{Jahn}, {Sales}, {Wetzel}, {Boylan-Kolchin},
  {Chan}, {El-Badry}, {Lazar}  \& {Bullock}}{{Jahn} et~al.}{2019}]{Jahn2019}
{Jahn} E.~D.,  {Sales} L.~V.,  {Wetzel} A.,  {Boylan-Kolchin} M.,  {Chan}
  T.~K.,  {El-Badry} K.,  {Lazar} A.,   {Bullock} J.~S.,  2019, \mn@doi
  [\mnras] {10.1093/mnras/stz2457}, \href
  {https://ui.adsabs.harvard.edu/abs/2019MNRAS.489.5348J} {489, 5348}

\bibitem[\protect\citeauthoryear{{Jenkins}}{{Jenkins}}{2013}]{Jenkins2013}
{Jenkins} A.,  2013, \mn@doi [\mnras] {10.1093/mnras/stt1154}, \href
  {http://adsabs.harvard.edu/abs/2013MNRAS.434.2094J} {434, 2094}

\bibitem[\protect\citeauthoryear{{Jethwa}, {Erkal}  \& {Belokurov}}{{Jethwa}
  et~al.}{2016}]{Jethwa2016}
{Jethwa} P.,  {Erkal} D.,   {Belokurov} V.,  2016, \mn@doi [\mnras]
  {10.1093/mnras/stw1343}, \href
  {https://ui.adsabs.harvard.edu/abs/2016MNRAS.461.2212J} {461, 2212}

\bibitem[\protect\citeauthoryear{{Kallivayalil}, {van der Marel}, {Besla},
  {Anderson}  \& {Alcock}}{{Kallivayalil} et~al.}{2013}]{Kallivayalil2013}
{Kallivayalil} N.,  {van der Marel} R.~P.,  {Besla} G.,  {Anderson} J.,
  {Alcock} C.,  2013, \mn@doi [\apj] {10.1088/0004-637X/764/2/161}, \href
  {https://ui.adsabs.harvard.edu/abs/2013ApJ...764..161K} {764, 161}

\bibitem[\protect\citeauthoryear{{Kallivayalil} et~al.,}{{Kallivayalil}
  et~al.}{2018}]{Kallivayalil2018}
{Kallivayalil} N.,  et~al., 2018, \mn@doi [\apj] {10.3847/1538-4357/aadfee},
  \href {https://ui.adsabs.harvard.edu/abs/2018ApJ...867...19K} {867, 19}

\bibitem[\protect\citeauthoryear{{Kim}, {Staveley-Smith}, {Dopita}, {Freeman},
  {Sault}, {Kesteven}  \& {McConnell}}{{Kim} et~al.}{1998}]{Kim1998}
{Kim} S.,  {Staveley-Smith} L.,  {Dopita} M.~A.,  {Freeman} K.~C.,  {Sault}
  R.~J.,  {Kesteven} M.~J.,   {McConnell} D.,  1998, \mn@doi [\apj]
  {10.1086/306030}, \href
  {https://ui.adsabs.harvard.edu/abs/1998ApJ...503..674K} {503, 674}

\bibitem[\protect\citeauthoryear{{Komatsu} et~al.,}{{Komatsu}
  et~al.}{2011}]{Komatsu2011}
{Komatsu} E.,  et~al., 2011, \mn@doi [\apjs] {10.1088/0067-0049/192/2/18},
  \href {http://adsabs.harvard.edu/abs/2011ApJS..192...18K} {192, 18}

\bibitem[\protect\citeauthoryear{{Koposov}, {Belokurov}, {Torrealba}  \&
  {Evans}}{{Koposov} et~al.}{2015}]{Koposov15a}
{Koposov} S.~E.,  {Belokurov} V.,  {Torrealba} G.,   {Evans} N.~W.,  2015,
  \mn@doi [\apj] {10.1088/0004-637X/805/2/130}, \href
  {https://ui.adsabs.harvard.edu/\#abs/2015ApJ...805..130K} {805, 130}

\bibitem[\protect\citeauthoryear{{Mao}, {Geha}, {Wechsler}, {Weiner},
  {Tollerud}, {Nadler}  \& {Kallivayalil}}{{Mao} et~al.}{2020}]{Mao2020}
{Mao} Y.-Y.,  {Geha} M.,  {Wechsler} R.~H.,  {Weiner} B.,  {Tollerud} E.~J.,
  {Nadler} E.~O.,   {Kallivayalil} N.,  2020, arXiv e-prints, \href
  {https://ui.adsabs.harvard.edu/abs/2020arXiv200812783M} {p. arXiv:2008.12783}

\bibitem[\protect\citeauthoryear{{McConnachie}}{{McConnachie}}{2012}]{McConnachie2012}
{McConnachie} A.~W.,  2012, \mn@doi [\aj] {10.1088/0004-6256/144/1/4}, \href
  {https://ui.adsabs.harvard.edu/abs/2012AJ....144....4M} {144, 4}

\bibitem[\protect\citeauthoryear{{McConnachie} \& {Venn}}{{McConnachie} \&
  {Venn}}{2020}]{McConnachie2020}
{McConnachie} A.~W.,  {Venn} K.~A.,  2020, \mn@doi [\aj]
  {10.3847/1538-3881/aba4ab}, \href
  {https://ui.adsabs.harvard.edu/abs/2020AJ....160..124M} {160, 124}

\bibitem[\protect\citeauthoryear{{McMillan}}{{McMillan}}{2011}]{McMillan2011}
{McMillan} P.~J.,  2011, \mn@doi [\mnras] {10.1111/j.1365-2966.2011.18564.x},
  \href {https://ui.adsabs.harvard.edu/abs/2011MNRAS.414.2446M} {414, 2446}

\bibitem[\protect\citeauthoryear{{Moster}, {Naab}  \& {White}}{{Moster}
  et~al.}{2013}]{Moster2013}
{Moster} B.~P.,  {Naab} T.,   {White} S.~D.~M.,  2013, \mn@doi [MNRAS]
  {10.1093/mnras/sts261}, \href
  {http://adsabs.harvard.edu/abs/2013MNRAS.428.3121M} {428, 3121}

\bibitem[\protect\citeauthoryear{{Nadler}, {Mao}, {Green}  \&
  {Wechsler}}{{Nadler} et~al.}{2019}]{Nadler2019}
{Nadler} E.~O.,  {Mao} Y.-Y.,  {Green} G.~M.,   {Wechsler} R.~H.,  2019,
  \mn@doi [\apj] {10.3847/1538-4357/ab040e}, \href
  {https://ui.adsabs.harvard.edu/abs/2019ApJ...873...34N} {873, 34}

\bibitem[\protect\citeauthoryear{{Nadler} et~al.,}{{Nadler}
  et~al.}{2020}]{Nadler2020}
{Nadler} E.~O.,  et~al., 2020, \mn@doi [\apj] {10.3847/1538-4357/ab846a}, \href
  {https://ui.adsabs.harvard.edu/abs/2020ApJ...893...48N} {893, 48}

\bibitem[\protect\citeauthoryear{{Newton}, {Cautun}, {Jenkins}, {Frenk}  \&
  {Helly}}{{Newton} et~al.}{2018}]{Newton2018}
{Newton} O.,  {Cautun} M.,  {Jenkins} A.,  {Frenk} C.~S.,   {Helly} J.~C.,
  2018, \mn@doi [\mnras] {10.1093/mnras/sty1085}, \href
  {https://ui.adsabs.harvard.edu/abs/2018MNRAS.479.2853N} {479, 2853}

\bibitem[\protect\citeauthoryear{{Pardy} et~al.,}{{Pardy}
  et~al.}{2020}]{Pardy2020}
{Pardy} S.~A.,  et~al., 2020, \mn@doi [\mnras] {10.1093/mnras/stz3192}, \href
  {https://ui.adsabs.harvard.edu/abs/2020MNRAS.492.1543P} {492, 1543}

\bibitem[\protect\citeauthoryear{{Patel}, {Besla}  \& {Sohn}}{{Patel}
  et~al.}{2017a}]{Patel2017}
{Patel} E.,  {Besla} G.,   {Sohn} S.~T.,  2017a, \mn@doi [\mnras]
  {10.1093/mnras/stw2616}, \href
  {https://ui.adsabs.harvard.edu/abs/2017MNRAS.464.3825P} {464, 3825}

\bibitem[\protect\citeauthoryear{{Patel}, {Besla}  \& {Mandel}}{{Patel}
  et~al.}{2017b}]{Patel2017b}
{Patel} E.,  {Besla} G.,   {Mandel} K.,  2017b, \mn@doi [\mnras]
  {10.1093/mnras/stx698}, \href
  {https://ui.adsabs.harvard.edu/abs/2017MNRAS.468.3428P} {468, 3428}

\bibitem[\protect\citeauthoryear{{Patel} et~al.,}{{Patel}
  et~al.}{2020}]{Patel2020}
{Patel} E.,  et~al., 2020, \mn@doi [\apj] {10.3847/1538-4357/ab7b75}, \href
  {https://ui.adsabs.harvard.edu/abs/2020ApJ...893..121P} {893, 121}

\bibitem[\protect\citeauthoryear{{Qu} et~al.,}{{Qu} et~al.}{2017}]{Qu2017}
{Qu} Y.,  et~al., 2017, \mn@doi [\mnras] {10.1093/mnras/stw2437}, \href
  {https://ui.adsabs.harvard.edu/abs/2017MNRAS.464.1659Q} {464, 1659}

\bibitem[\protect\citeauthoryear{{Sales}, {Navarro}, {Abadi}  \&
  {Steinmetz}}{{Sales} et~al.}{2007}]{Sales2007}
{Sales} L.~V.,  {Navarro} J.~F.,  {Abadi} M.~G.,   {Steinmetz} M.,  2007,
  \mn@doi [\mnras] {10.1111/j.1365-2966.2007.12026.x}, \href
  {https://ui.adsabs.harvard.edu/abs/2007MNRAS.379.1475S} {379, 1475}

\bibitem[\protect\citeauthoryear{{Sales}, {Navarro}, {Cooper}, {White}, {Frenk}
   \& {Helmi}}{{Sales} et~al.}{2011}]{Sales2011}
{Sales} L.~V.,  {Navarro} J.~F.,  {Cooper} A.~P.,  {White} S. D.~M.,  {Frenk}
  C.~S.,   {Helmi} A.,  2011, \mn@doi [\mnras]
  {10.1111/j.1365-2966.2011.19514.x}, \href
  {https://ui.adsabs.harvard.edu/abs/2011MNRAS.418..648S} {418, 648}

\bibitem[\protect\citeauthoryear{{Sales}, {Wang}, {White}  \&
  {Navarro}}{{Sales} et~al.}{2013}]{Sales2013}
{Sales} L.~V.,  {Wang} W.,  {White} S. D.~M.,   {Navarro} J.~F.,  2013, \mn@doi
  [\mnras] {10.1093/mnras/sts054}, \href
  {https://ui.adsabs.harvard.edu/abs/2013MNRAS.428..573S} {428, 573}

\bibitem[\protect\citeauthoryear{{Sales} et~al.,}{{Sales}
  et~al.}{2017}]{Sales2017}
{Sales} L.~V.,  et~al., 2017, \mn@doi [\mnras] {10.1093/mnras/stw2461}, \href
  {http://adsabs.harvard.edu/abs/2017MNRAS.464.2419S} {464, 2419}

\bibitem[\protect\citeauthoryear{{Samuel} et~al.,}{{Samuel}
  et~al.}{2020}]{Samuel2020}
{Samuel} J.,  et~al., 2020, \mn@doi [\mnras] {10.1093/mnras/stz3054}, \href
  {https://ui.adsabs.harvard.edu/abs/2020MNRAS.491.1471S} {491, 1471}

\bibitem[\protect\citeauthoryear{{Sawala} et~al.,}{{Sawala}
  et~al.}{2016}]{Sawala2016}
{Sawala} T.,  et~al., 2016, \mn@doi [\mnras] {10.1093/mnras/stw145}, \href
  {https://ui.adsabs.harvard.edu/abs/2016MNRAS.457.1931S} {457, 1931}

\bibitem[\protect\citeauthoryear{{Schaye} et~al.,}{{Schaye}
  et~al.}{2015}]{Schaye2015}
{Schaye} J.,  et~al., 2015, \mn@doi [\mnras] {10.1093/mnras/stu2058}, \href
  {http://adsabs.harvard.edu/abs/2015MNRAS.446..521S} {446, 521}

\bibitem[\protect\citeauthoryear{{Sch{\"o}nrich}, {Binney}  \&
  {Dehnen}}{{Sch{\"o}nrich} et~al.}{2010}]{Schonrich2010}
{Sch{\"o}nrich} R.,  {Binney} J.,   {Dehnen} W.,  2010, \mn@doi [\mnras]
  {10.1111/j.1365-2966.2010.16253.x}, \href
  {https://ui.adsabs.harvard.edu/abs/2010MNRAS.403.1829S} {403, 1829}

\bibitem[\protect\citeauthoryear{{Shao}, {Cautun}, {Deason}, {Frenk}  \&
  {Theuns}}{{Shao} et~al.}{2018}]{Shao2018}
{Shao} S.,  {Cautun} M.,  {Deason} A.~J.,  {Frenk} C.~S.,   {Theuns} T.,  2018,
  \mn@doi [\mnras] {10.1093/mnras/sty1470}, \href
  {https://ui.adsabs.harvard.edu/abs/2018MNRAS.479..284S} {479, 284}

\bibitem[\protect\citeauthoryear{{Springel}}{{Springel}}{2005}]{Springel2005b}
{Springel} V.,  2005, \mn@doi [\mnras] {10.1111/j.1365-2966.2005.09655.x},
  \href
  {http://esoads.eso.org/cgi-bin/nph-bib_query?bibcode=2005MNRAS.364.1105S&db_key=AST}
  {364, 1105}

\bibitem[\protect\citeauthoryear{{Springel}, {Yoshida}  \& {White}}{{Springel}
  et~al.}{2001}]{Springel2001}
{Springel} V.,  {Yoshida} N.,   {White} S. D.~M.,  2001, \mn@doi [\na]
  {10.1016/S1384-1076(01)00042-2}, \href
  {https://ui.adsabs.harvard.edu/abs/2001NewA....6...79S} {6, 79}

\bibitem[\protect\citeauthoryear{{Tollerud}, {Boylan-Kolchin}, {Barton},
  {Bullock}  \& {Trinh}}{{Tollerud} et~al.}{2011}]{Tollerud2011}
{Tollerud} E.~J.,  {Boylan-Kolchin} M.,  {Barton} E.~J.,  {Bullock} J.~S.,
  {Trinh} C.~Q.,  2011, \mn@doi [\apj] {10.1088/0004-637X/738/1/102}, \href
  {https://ui.adsabs.harvard.edu/abs/2011ApJ...738..102T} {738, 102}

\bibitem[\protect\citeauthoryear{{Wang}, {Frenk}, {Navarro}, {Gao}  \&
  {Sawala}}{{Wang} et~al.}{2012}]{Wang2012}
{Wang} J.,  {Frenk} C.~S.,  {Navarro} J.~F.,  {Gao} L.,   {Sawala} T.,  2012,
  \mn@doi [\mnras] {10.1111/j.1365-2966.2012.21357.x}, \href
  {https://ui.adsabs.harvard.edu/abs/2012MNRAS.424.2715W} {424, 2715}

\bibitem[\protect\citeauthoryear{{Westerlund}}{{Westerlund}}{1990}]{Westerlund1990}
{Westerlund} B.~E.,  1990, \mn@doi [\aapr] {10.1007/BF00873541}, \href
  {https://ui.adsabs.harvard.edu/abs/1990A&ARv...2...29W} {2, 29}

\bibitem[\protect\citeauthoryear{{Woo}, {Courteau}  \& {Dekel}}{{Woo}
  et~al.}{2008}]{Woo2008}
{Woo} J.,  {Courteau} S.,   {Dekel} A.,  2008, \mn@doi [\mnras]
  {10.1111/j.1365-2966.2008.13770.x}, \href
  {https://ui.adsabs.harvard.edu/abs/2008MNRAS.390.1453W} {390, 1453}

\bibitem[\protect\citeauthoryear{{Yozin} \& {Bekki}}{{Yozin} \&
  {Bekki}}{2015}]{Yozin2015}
{Yozin} C.,  {Bekki} K.,  2015, \mn@doi [\mnras] {10.1093/mnras/stv1828}, \href
  {https://ui.adsabs.harvard.edu/abs/2015MNRAS.453.2302Y} {453, 2302}

\bibitem[\protect\citeauthoryear{{van der Marel} \& {Kallivayalil}}{{van der
  Marel} \& {Kallivayalil}}{2014}]{vanderMarel2014}
{van der Marel} R.~P.,  {Kallivayalil} N.,  2014, \mn@doi [\apj]
  {10.1088/0004-637X/781/2/121}, \href
  {https://ui.adsabs.harvard.edu/abs/2014ApJ...781..121V} {781, 121}

\bibitem[\protect\citeauthoryear{{van der Marel}, {Alves}, {Hardy}  \&
  {Suntzeff}}{{van der Marel} et~al.}{2002}]{vanderMarel2002}
{van der Marel} R.~P.,  {Alves} D.~R.,  {Hardy} E.,   {Suntzeff} N.~B.,  2002,
  \mn@doi [\aj] {10.1086/343775}, \href
  {https://ui.adsabs.harvard.edu/abs/2002AJ....124.2639V} {124, 2639}

\makeatother
\end{thebibliography}

\end{document}